\documentclass{ieeeaccess}

\usepackage{tikz}
\usetikzlibrary{quantikz2}
\NewSpotColorSpace{PANTONE}
\AddSpotColor{PANTONE} {PANTONE3015C} {PANTONE\SpotSpace 3015\SpotSpace C} {1 0.3 0 0.2}
\SetPageColorSpace{PANTONE}%

\PassOptionsToPackage{hyphens}{url}\usepackage{hyperref}
\usepackage{amsmath, amssymb, amsfonts}
\usepackage{graphicx}
\usepackage{textcomp}
\usepackage{caption}
\usepackage{comment}
\usepackage{algorithm}
\usepackage[noend]{algpseudocode}
\usepackage{booktabs}
\usepackage{mathtools}
\usepackage{subfigure}
\usepackage{tabularx}
\usepackage{makecell}
\usepackage{multirow}
\usepackage{cite}

\begin{document}

\history{Date of publication xxxx 00, 0000, date of current version xxxx 00, 0000.}
\doi{10.1109/TQE.2020.DOI}

\title{Hardware-aware and Resource-efficient Circuit Packing and Scheduling on Trapped-Ion Quantum Computers}

\author{\uppercase{Miguel Palma}\authorrefmark{1},
\uppercase{Shuwen Kan} \authorrefmark{1},
\uppercase{Wenqi Wei} \authorrefmark{1},
\uppercase{Juntao Chen} \authorrefmark{1},
\uppercase{Kaixun Hua} \authorrefmark{2},
\uppercase{Sara Mouradian} \authorrefmark{3},
\uppercase{and Ying Mao}\authorrefmark{1},
}
\address[1]{Fordham University, Bronx, NY 10458 USA}
\address[2]{University of South Florida, Tampa, FL 33620 USA}
\address[3]{University of Washington, Seattle, WA 98195 USA}

\tfootnote{This research was supported in part by the National Science Foundation (NSF) under grant agreements 2301884, 2329020, 2335788, and 2343535. 
}

\markboth
{Palma \headeretal: Hardware-aware and Resource-efficient Circuit Packing and Scheduling on Trapped-Ion Quantum Computers}
{Palma \headeretal: Hardware-aware and Resource-efficient Circuit Packing and Scheduling on Trapped-Ion Quantum Computers}

\corresp{Corresponding author: Miguel Palma (email: mip2@fordham.edu) and Ying Mao (ymao41@fordham.edu).}



\begin{abstract}
The rapid expansion of quantum cloud services has led to long job queues due to single-tenant execution models that underutilize hardware resources. Quantum multi-programming (QMP) mitigates this by executing multiple circuits in parallel on a single device, but existing methods target superconducting systems with limited connectivity, high crosstalk, and lower gate fidelity. Trapped-ion architectures, with all-to-all connectivity, long coherence times, and high-fidelity mid-circuit measurement properties, presents itself as a more suitable platform for scalable QMP.
We present CircPack, a hardware-aware circuit packing framework designed for modular trapped-ion devices based on the Quantum Charge-Coupled Device (QCCD) architecture. CircPack formulates static circuit scheduling as a two-dimensional packing problem with hardware-specific shuttling constraints. Compared to superconducting-based QMP approaches, CircPack achieves up to 70.72\% better fidelity, 62.67\% higher utilization, and 32.80\% improved layer reduction.
This framework is also capable of scalable, balanced scheduling across a cluster of independent QCCD modules, highlighting trapped-ion systems’ potential in improving the throughput of quantum cloud computing in the near future.

\end{abstract}


\titlepgskip=-15pt

\maketitle

\section{Introduction}

The development and maintenance of advanced quantum computer systems present formidable challenges~\cite{almudever2017engineering, de2021materials, mohseni2024build}, limiting physical access to a small number of researchers in academia, industry, and national laboratories. 
With the rise of distributed quantum computing~\cite{main2025distributed, caleffi2024distributed}—including advances in workload management~\cite{mantha2025pilot, kan2024scalable, kan2024circuit, nguyen2024iquantum, du2024efficient}, distributed compilers~\cite{ferrari2023modular, du2024hardware, chatterjee2022qurzon, cuomo2023optimized, du2025hardware}, and resource-efficient distributed applications~\cite{jiang2024resource, robert2021resource, du2026distributed}, the need for quantum cloud computing has become increasingly critical.
Quantum cloud computing, also known as Quantum as a Service (QaaS), has emerged to address this accessibility barrier~\cite{golec2024quantum} by enabling remote access to quantum hardware via the internet. This paradigm eliminates the necessity for physical proximity to quantum devices, offers a flexible pay-as-you-go model, and relieves users from the burden of hardware operation and maintenance. Leading providers such as IBM Quantum, Azure Quantum, and Amazon Braket now offer access to a diverse range of quantum processing units (QPUs), each featuring distinct characteristics such as qubit count, coupling topology, and gate error rates~\cite{ibmq_resources, azure_quantum_providers, amazon_braket_quantum_computers}.


The use of quantum cloud computing has grown rapidly. IBM Quantum cloud usage analysis revealed a 7$\times$ increase in job executions from 2019 to 2021, causing significant queueing delays \cite{ravi2021quantum}.  Median queue times reached 60 minutes, with 30\% of jobs waiting over two hours and 10\% exceeding a full day. More recently, D-Wave announced that their Q3 2024 quantum cloud revenue grew by 41\% year over year \cite{d_wave_revenue} and the overall market is forecasted to grow at a Compound Annual Growth Rate (CAGR) of 38\% from 2024 to 2031 \cite{quantum_market_forecast}. If hardware availability fails to keep up with demand, QaaS costs could increase in the future.

One of the root causes of this delay is the single-tenant nature of the current quantum cloud. Only one job at a time is allowed to run on a piece of quantum hardware to avoid any interference that could degrade the output fidelity. This means that a 150-qubit device is considered fully occupied by a 30-qubit circuit for its entire job duration, even if only one-fifth of the total qubits are needed.
Users can select QPUs with better calibration results
for execution, but this leads to an unbalanced job distribution across the cluster. To mitigate this impending bottleneck, improved resource utilization will be needed at both the individual device and system-wide levels.

Quantum multi-programming (QMP) offers a potential solution by executing multiple quantum circuits on a single device simultaneously, making use of idle qubits \cite{das2019case}. The spatial concurrency unlocked by QMP enables multi-tenancy, allowing quantum cloud clusters to scale out more effectively to service the demands of multiple users at the same time. Despite these promised benefits, QMP is not being utilized by quantum cloud providers today because obtaining high fidelity results is the top priority.

The superconducting circuit architecture is currently among the most widely used type of quantum devices. It boasts fast gate operation times but also exhibits limited 2-qubit gate connectivity, vulnerability to crosstalk noise, short decoherence times, and requires frequent device calibrations \cite{linke2017experimental}. Because of the heterogenous nature of its hardware resources, superconducting-based QMP approaches are unable to fully utilize all available qubits and suffers from significant reductions in output fidelity as well \cite{das2019case, ohkura2022simultaneous, liu2024qucloud+, niu2023enabling}. 
On the other hand, trapped-ion devices use low-maintenance natural qubits and offer full, mutable qubit connectivity. These properties allow it to perform complex circuits with high-fidelity, and enable qubit reuse via mid-circuit measurement and reset capabilities \cite{bruzewicz2019trapped}. 
Trapped-ion devices have traditionally been limited by slower gate speeds and smaller qubit capacities, but these scalability issues have been addressed. The Quantum Charge-Coupled Device (QCCD) architecture utilizes multiple small ion traps are connected with shuttling paths for movement to enable modularity\cite{pino2020demonstration}.

We introduce \textit{Circuit Packing}, a QMP framework designed to maximize hardware utilization and circuit throughput on QCCD trapped-ion architectures. The special characteristics of these devices allow us to treat the hardware qubits as homogenous virtual resources and model the QMP task as a constrained rectangle packing problem (RPP). Using this framework, improved throughput is achieved by reducing the number of circuit layers via parallel execution while remaining within the device's qubit capacity. The scheduling task can be applied for static circuits on both single-worker and multi-worker QPU systems, making it suitable for quantum cloud operations. Furthermore, we introduce hardware-aware optimizations that aim to mitigate fidelity loss, accepting slight reductions in packing efficiency in exchange for better quality results. 

The key contributions are as follows:
\begin{itemize}
    \item We define new metrics for quantum resource utilization, capturing both the spatial and temporal aspects of qubit allocation.
    \item We develop a novel problem formulation for quantum circuit scheduling on QCCD trapped-ion devices, incorporating the characteristics of the hardware. 
    \item We introduce CircPack, a rectangle-packing-based approach that generates circuit schedules with significantly improved resource utilization and throughput, while maintaining high fidelity through hardware-aware packing improvements. Using the H2 emulator \cite{quantinuum_h2_emu}, QCCDSim tool \cite{murali2020architecting}, and the MTS-QCCD compiler \cite{saki2022muzzle}, our results on a queue of 200 circuits achieves average utilization values of 89.35\%, circuit layer reduction of 77.35\%, and fidelity of 93.66\%.
    \item A comparative evaluation between CircPack and other previous QMP methods was performed, showing relative improvements of up to 70.72\% in fidelity, 62.07\% in utilization, and 32.80\% in LRF. These results present significant improvements in both fidelity and throughput of multi-programmed circuits.
    \item We demonstrate the suitability of CircPack to perform balanced distributed scheduling across multiple quantum workers using a simulated scenario with 5 workers and 3000 circuits. A minimal 2.83\% difference in makespan and 1.74\% for average utilization is observed between the workers with the highest and lowest loads. 
\end{itemize}

\section{Background and Related Work}
\subsection{Quantum Computing Basics}
The fundamental unit of quantum computation is called the quantum bit, also known as the \textit{qubit} \cite{vedral1998basics}.  Unlike silicon-based classical bits, there are many physical qubit implementations, with their own different strengths and weaknesses. Superconducting circuits are currently the dominant qubit technology; however, trapped ions, neutral atoms, and other approaches are actively being pursued towards fault-tolerant computation \cite{bhat2022quantum}.


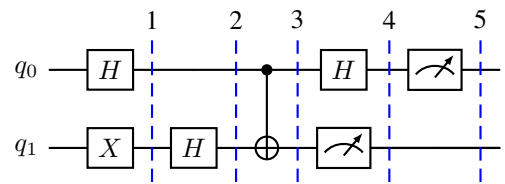
\begin{figure} [H]
    \centering
    \begin{quantikz} [slice all, slice style=blue]
        \lstick{$q_0$} & \gate{H} & & \ctrl{1} & \gate{H} & \meter{} & \\
        \lstick{$q_1$} & \gate{X} & \gate{H} & \targ{} & \meter{} & \qw & \qw
    \end{quantikz}
    \caption{A quantum circuit with a width of 2 qubits and depth of 5 layers.}
    \label{fig:example_circ}
\end{figure}

The standard abstraction in quantum computing is known as the \textit{quantum circuit model} \cite{deutsch1989quantum, nielsen2010quantum}, illustrated in Figure \ref{fig:example_circ}. In here, each qubit is represented by a "wire" that begins in a defined state and undergoes a sequence of quantum gates, which are drawn as blocks. These operations can be grouped into concurrent sets known as \textit{layers}, which are denoted here with the dashed blue lines.
A circuit's \textit{width} refers to the number of qubits in use while the \textit{depth} refers to the total number of layers in the circuit. 
Together, these define the space and time requirements for any given circuit.

\begin{figure}
    \centering
    \includegraphics[width=0.90\linewidth]{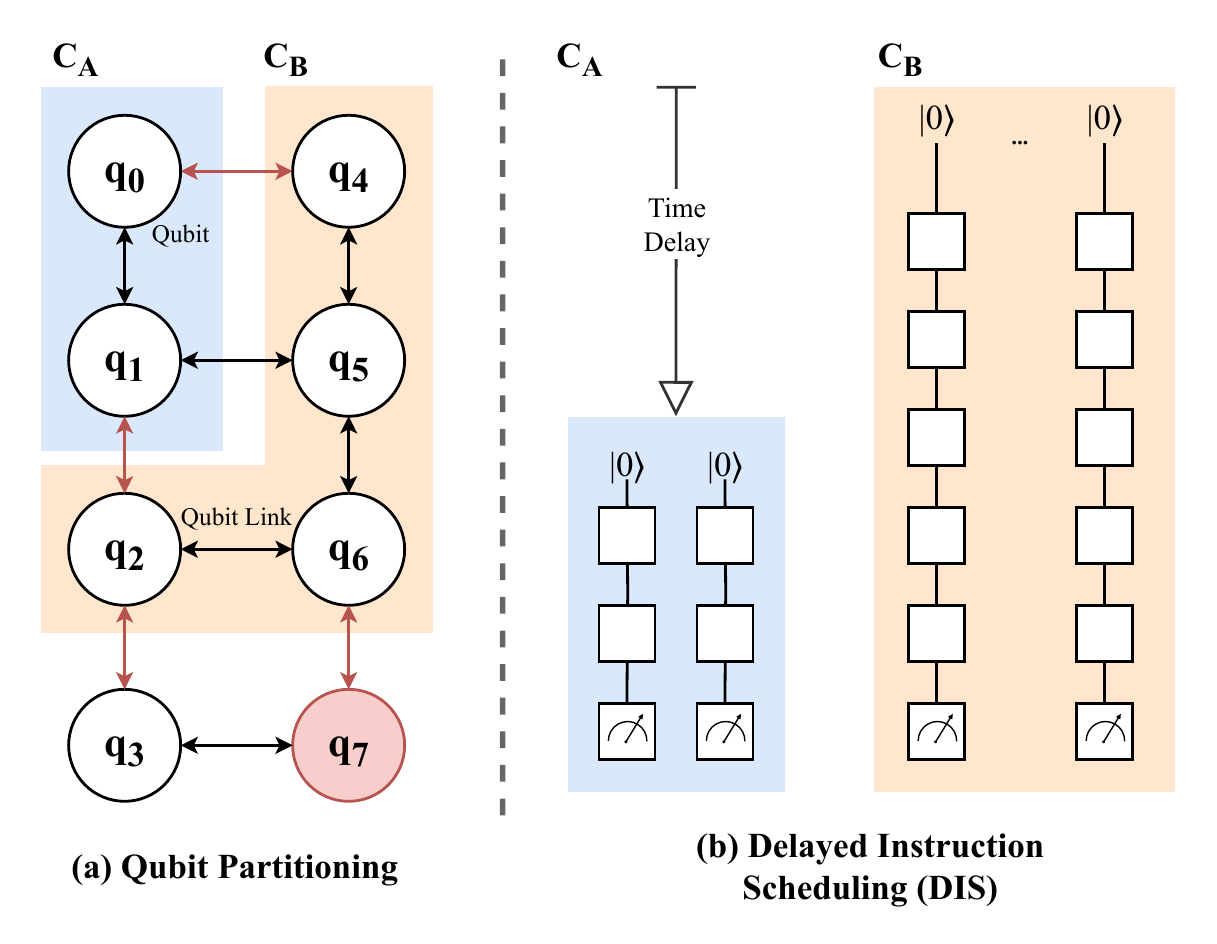}

    \caption{Quantum Multi-programming as performed on Superconducting circuit architectures (e.g., IBM-Q).}
    \label{fig:sc_multiprog}
\end{figure}

\subsection{Quantum Multi-programming}




Quantum multi-programming (QMP) refers to single-device quantum circuit concurrency and is applicable in cases where the hardware has a higher qubit capacity than the total qubit requirements of the input circuits. Previous work on QMP chose superconducting circuit devices because of their popularity and high qubit count but the results were limited by the hardware \cite{das2019case, liu2021qucloud, ohkura2022simultaneous, niu2023enabling}.

Superconducting devices are fundamentally constrained to its coupling graph, which encapsulates its qubit connectivity and error profiles. The primary goal of superconducting-based QMP is identifying qubit subsets that can handle reliable independent circuit executions, a non-trivial task that scales on the size of the coupling graph \cite{das2019case}. The variance in qubit calibration noise is considered when creating the partitions, making a choice between under-utilizing the device qubits or decreasing the output quality. Even then, the resulting partition may have poor compatibility with the circuit's qubit interaction graph, requiring additional SWAP gates for routing. Figure \ref{fig:sc_multiprog} (a) shows the coupling graph of an example superconducting device where the red circle represents a qubit with noisy calibration results while the red arrows are qubit connections with high error rates.
    

On current Noisy Intermediate-Scale Quantum (NISQ) superconducting devices, measurement operations can introduce destructive interference, potentially corrupting the state of neighboring qubits \cite{sarovar2020detecting}. Delayed Instruction Scheduling (DIS) addresses this by postponing shorter circuits so that they are measured at the same time \cite{das2019case}, as shown in Figure \ref{fig:sc_multiprog} (b). Although the interference is avoided, qubits reserved for $C_A$ remain idle for the delay period. In cases where the circuits have a large difference in depth, this results in significant qubit time under-utilization.

Quantum multi-programming on trapped-ion systems is less studied but shows potential. The full connectivity of the device qubits simplifies the qubit assignment task and the compiler takes care of the underlying shuttle schedule.  
A benchmark study comparing QMP performance between superconducting and trapped-ion systems found significant differences in fidelity, ranging from 5-50\%.\cite{niu2022multi}. Compared to independent execution,  multi-programmed circuit fidelity drops by 3.4\% on superconducting while a decrease of only 0.5\% was observed on trapped-ion devices. The limitation of this study is that it only studied 2 small circuits running in parallel. 


\begin{figure*}
    \centering
    \includegraphics[width=0.9\linewidth]{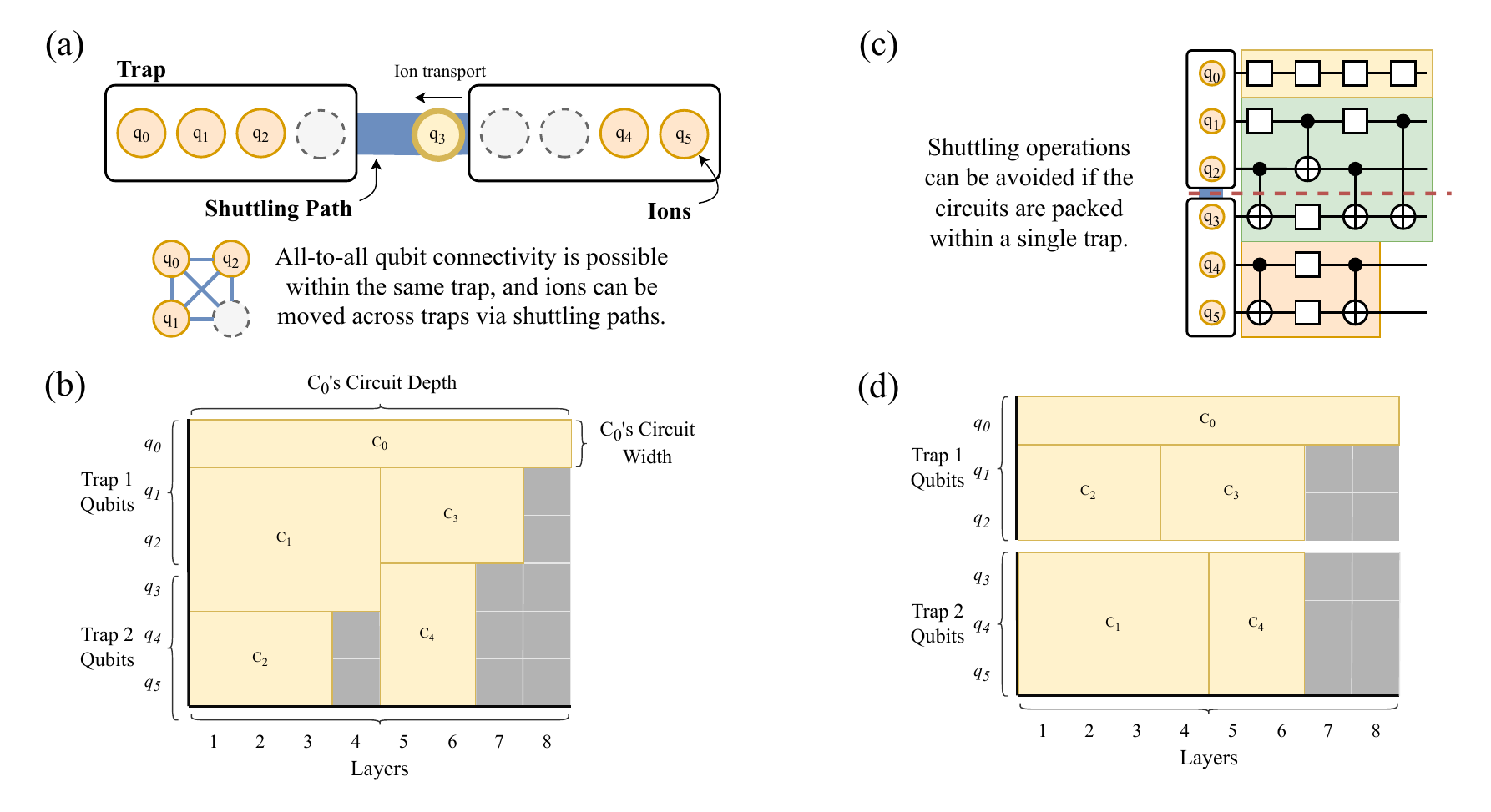}
    \caption{(a) QCCD trapped ion modular system topology. (b) Gantt chart representation of a circuit schedule. (c) Example combined circuit with 2-qubit gates across 2 separate traps. (d) Trap-based packing method avoids shuttle operations.}
    \label{fig:qccd architecture}
\end{figure*}

\subsection{Trapped-Ion Quantum Computers and QCCD Architecture}

Trapped-ion quantum computers are a leading candidate for scalable quantum computation due to their intrinsic physical advantages \cite{bruzewicz2019trapped}. Individual ions are confined within electromagnetic fields to serve as qubits, 
and gate operations are performed using laser pulses. Trapped-ion systems are characterized by long coherence times, low crosstalk, and support for high fidelity mid-circuit measurement and reset (MCMR), enabling qubit reuse in complex quantum circuits \cite{decross2023qubit}.

To achieve scalability, modern trapped-ion devices adopt the Quantum Charge-Coupled Device(QCCD) architecture \cite{pino2020demonstration}, which introduces a modular layout comprising distinct \textit{storage zones} and \textit{gate zones}. As illustrated in Figure~\ref{fig:qccd architecture} (a), ions can be shuttled to different zones as needed via shuttling paths and junctions. 

This modularity enables extensibility by simply adding more traps or interaction zones, providing a clear path toward larger quantum processors. Furthermore, the ability to move qubits between zones effectively realizes all-to-all connectivity—overcoming the static coupling constraints seen in superconducting circuits.

However, this flexibility comes with trade-offs. The act of shuttling ions introduces mechanical motion, requires frequent cooling operations, and increases total operation time-- all of these directly contributing to decreased result fidelity. For instance, on the Quantinuum H2-1 system, only 1--2\% of execution time is devoted to quantum gates, while over 60\% is consumed by ion transport and an additional 38\% by cooling procedures \cite{moses2023race}. These overheads highlight the dominant role of shuttling in the performance and noise profile of QCCD-based systems.

\subsection{Related Work}

Quantum multi-programming (QMP) has been proposed as a solution to improve resource utilization and reduce execution latency on quantum devices. From a scheduling and mapping perspective, several QMP methods have been proposed for superconducting architectures. 
QuMC~\cite{niu2023enabling} introduces a compiler that partitions qubits into crosstalk-aware regions using randomized benchmarking and optimized mapping to enable simultaneous circuit execution. 
QuCloud+~\cite{liu2024qucloud+} presents a fidelity-driven mapping framework that supports
2D/3D superconducting topologies, combining crosstalk-aware community detection, qubit-degree allocation, and 
a mechanism to reduce SWAP overhead.
Palloq~\cite{ohkura2022simultaneous} offers a comprehensive runtime and fidelity analysis for QMP using buffer qubits.


For trapped-ion systems, Quantinuum published a Circuit Stitching workflow in their documentation~\cite{quantinuum_circuit_stitching} that merges multiple circuits into a single execution batch using MCMR. It requires all input circuits to be of equal width but it maximizes the qubit utilization by leveraging the full qubit connectivity capability.
Although job overhead is reduced for the circuit executions, it uses a FIFO-based policy without optimizing for spatial or temporal resource utilization, nor does it account for circuit depths during the co-scheduling.



\subsection{Motivation}

Quantum multi-programming is a viable near-term solution to address increased queueing times on the quantum cloud. As the number of qubits per device continues to scale, adopting a multi-tenant execution environment has the potential to improve hardware resource utilization.
However, prior work in both superconducting and trapped-ion QMP achieve a limited utilization of spatial qubit resources and, more importantly, overlook the temporal aspect of parallel execution. Circuits are processed for co-execution in discrete batches, selected in a first-come first-serve basis without regard to the difference in circuit depth. Our approach, enabled by the hardware characteristics of trapped-ion systems, performs multiprogramming in a continuous manner to fully utilize available qubit space and time resources, minimize total execution layers, while maintaining high-fidelity output.

\section{Solution Design}

\subsection{Circuit Packing Framework}
In this section, we lay out the theoretical foundations of our framework. Here are our assumptions on the type and constraints of the input circuits:
\begin{itemize}
    \item The input circuits have a fixed width and depth after compilation.
    \item To simplify the model, we assume that all gate operation times are equivalent.
    \item Circuits may be executed in any order, but a best effort approach is applied to maintain a reasonable wait time for users.
\end{itemize}

Dynamic circuits \cite{corcoles2021exploiting} leverage mid-circuit measurement to perform adaptive gate sequences based on intermediate results, which are used in quantum error correction (QEC) \cite{devitt2013quantum} and qubit-reuse \cite{decross2023qubit} schemes. However, these variable operations cannot be directly integrated into the circuit packing framework, as their execution depths may vary across runs unless a maximum value is predetermined. To address this, our approach introduces a dynamic window-based cutoff mechanism that accommodates circuits with differing depths within a defined window. The window size can be set to the maximum depth of the dynamic circuits at the compilation level, ensuring compatibility with the circuit packing framework.

In our framework, we have a key assumption that all gate operations have equal time because the packing is performed on the circuit layer level while the compiler produces a schedule for shuttling and gate operations. In QCCD systems, these shuttling operations have been demonstrated to take up the largest share of the total execution time \cite{moses2023race}. Our trap-aware packing method, discussed later in this section, attempts to mitigate this by reducing the number of shuttling operations needed.
 

\subsubsection{Quantum Circuit Scheduling}



The main output of a QMP algorithm is a \textit{circuit schedule} that details the exact circuit execution order and qubit allocation. An ideal schedule maximizes resource utilization, while maintaining an acceptable result quality. 

Gantt charts are used in logistics planning to visualize task schedules that show resource assignment, job durations, and dependencies \cite{wilson2003gantt}. In a circuit schedule Gantt chart, such as Figure \ref{fig:qccd architecture} (b), the y-axis represents the qubits in the device and the x-axis is the number of layers or time dimension. Each rectangle in the diagram is a quantum circuit where its height corresponds to the circuit width, and the rectangle width corresponds to the circuit depth. The position of each rectangle in the diagram is a mapping between a circuit to a subset of the device qubits for a given time interval.

The hardware characteristics of trapped-ion systems map neatly into the Gantt chart representation. The inherent all-to-all qubit connectivity of these systems \cite{moses2023race} enables qubit assignment without needing to consider the SWAP costs while low crosstalk MCMR enables qubit reuse and as-early-as-possible scheduling \cite{decross2023qubit}. Instead of leaving qubits idle, we can schedule shorter circuits along the depth of a long circuit to maximize the temporal utilization of the device qubits. Given these properties, our main interest is to determine an optimal method to schedule a large set of circuits in the Gantt chart representation.

\subsubsection{Rectangle Packing}

The 2D rectangle packing problem (RPP) is an NP-hard combinatorial optimization task traditionally used in pallet loading and job shop scheduling. Its objective is to determine the minimum-sized rectangular container needed to pack a given set of rectangles, $R = \{r_0, ... , r_n\}$, such that none of them overlap. Each rectangle $r_i$ has a corresponding width $w_i$ and height $h_i$ while the container has a width $W$ and height $H$. For our framework, we restrict the container height to equal the qubit count and aim to minimize the container width (x-axis) which corresponds to the total number of layers to execute all the circuits in the schedule.


Due to its inherent complexity, this problem is approached using two primary methods: exact techniques, such as constraint programming, which find optimal solutions but take much longer to compute, and heuristic-based algorithms, which sacrifice optimality to provide approximate solutions faster. A comprehensive survey \cite{jylanki2010thousand} evaluates numerous heuristic methods suitable for online applications. Among these, the Skyline algorithm stands out for its balance of performance and simplicity. By tracking only the upper boundary of placed rectangles, it achieves a time complexity of $O(n^2)$ and a space complexity of $O(n)$.

\subsection{Hardware-aware Packing}
Reducing the number of shuttling operations is a key factor to improving the result fidelity on QCCD architectures\cite{murali2020architecting, saki2022muzzle}. By taking the trap configuration and device characteristics into account during the packing process, we propose two methods that aim to mitigate noise in the produced circuit schedule.

\subsubsection{Trap-based Packing}
\label{section:trap_based_packing}

A method of reducing the number of shuttles needed for a circuit schedule is to integrate the trap layout information in the circuit packing process. As seen in Figure \ref{fig:qccd architecture} (c), we propose minimizing the number of circuit layers that are assigned across multiple traps. By modeling each trap as a separate instance of the rectangle packing problem, we ensure that each of the circuits contained do not incur any shuttles (Figure \ref{fig:qccd architecture} (d)). A 6-qubit device comprising two traps connected by a single junction, can be reduced to performing packing across two smaller containers of height 3. However, this simplification may reduce overall packing efficiency compared to a single, undivided container.

\begin{figure}
    \centering
    \includegraphics[width=0.85\linewidth]{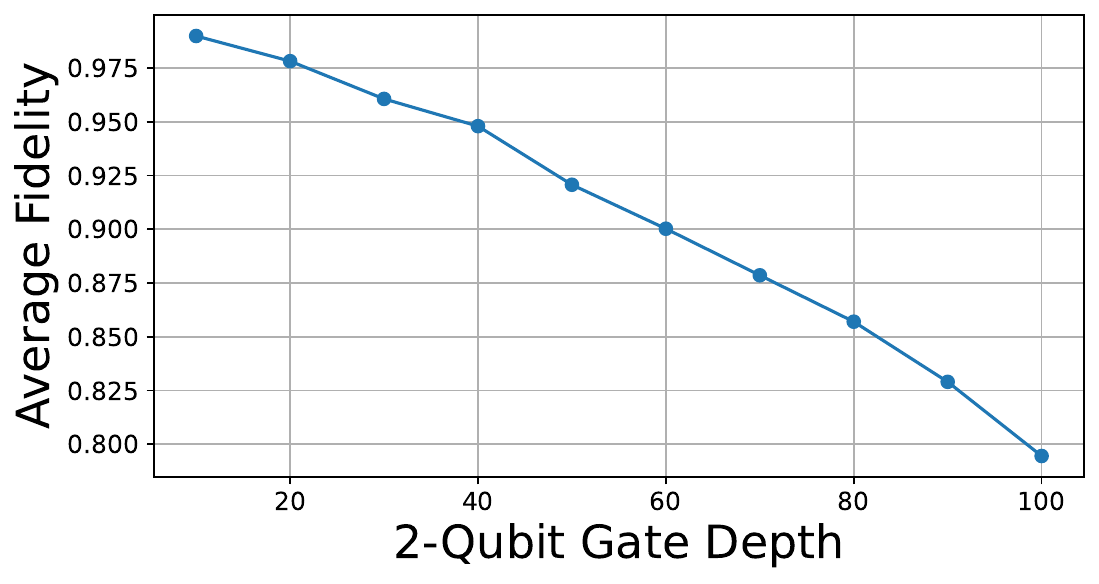}
    \caption{Fidelity vs 2-Qubit gate depth on random circuits using the MTS-QCCD and QCCDSim \cite{saki2022muzzle, murali2020architecting}.}
    \label{fig:cnot_depth_graph}
\end{figure}

\subsubsection{2-Qubit Gate-based Scheduling Cutoff} \label{section:2qcutoff}
One drawback of this framework is that it continuously packs circuits into the Gantt chart until you run out of items in the queue.
For large circuit sets, this results in schedules with extremely long makespans and may require many ion shuttles. Extended device operation also prevents the system from resetting to release the accumulated heat and motion via cooling. To address this, we introduce a windowing approach to divide a schedule into multiple batches when the running count of two-qubit gates exceeds a parameter, $\alpha$.

It is important to determine an optimal value for the scheduling cutoff parameter $\alpha$. The value must be large enough to allow for reasonable throughput while ensuring that output fidelity remains within acceptable bounds. For this, we conducted an experiment using the MST-QCCD-Compiler \cite{saki2022muzzle} based on the QCCDSim \cite{murali2020architecting} to generate a shuttling schedule and estimate fidelity using the number of ion transport operations using a linear 2-trap topology with 10 qubits each.
We filled 30 circuits with random CNOT pairs until the 2-qubit gate depth reaches $D, D \in [10, 20, \dots, 100]$.  As shown in Figure \ref{fig:cnot_depth_graph}, the average fidelity decreases nearly linearly as the 2-qubit gate depth increases. To target a minimum fidelity value of around 90\%, we set the value of $\alpha$ to be 170, which is the average number of 2Q gates in our random circuits when the 2Q gate depth is 60.

Implementing the cutoff offers an additional benefit: faster user feedback.  Consider a queue of 30 jobs, each requiring 1000 shots, scheduled on a single device. Without batching, users must wait for the entire schedule to complete before receiving any results. Although system-wide throughput is optimized, the individual user experience greatly suffers. By partitioning a schedule into batches, we balance throughput improvements with an acceptable Quality of Service (QoS).



\begin{figure*}
    \centering
    \includegraphics[width=0.95\linewidth]{images/pdfs/system_overview.pdf}
    \caption{An overview of the system comprised of the Circuit Scheduler and Execution Handler, along with the detailed Inputs and Outputs. First, (1) the circuits are preprocessed via native-gate compilation which are then (2) processed by the packing algorithm. Using the schedule generated from the packing, (3) the subcircuits are combined into a single QASM to be (4) sent for execution in the QPUs. Finally, (5) the result bitstrings are unbundled and the individual subcircuit fidelity results can be obtained.}
    \label{fig:process_diagram}
\end{figure*}

\subsection{System Design}
Figure \ref{fig:process_diagram} shows the system diagram for the Circuit Packing framework,
including job assignment across distributed QPUs. The main components, the quantum cloud job queue, the circuit scheduler, and the execution handler, operate in a continuous fashion similar to classical cloud systems. 

The central component in this system is the Circuit Scheduler which is responsible for the multi-programming mechanism, using the qubit count and trap topology hardware information of each QPU. Upon receiving the job queue, the individual circuits are first precompiled to the native gate level before the packings are determined. Our tests show that this method improved the result fidelities compared to running the compiler on the combined circuit. The updated circuit widths, depths, and 2-qubit gate counts are fed as the input for the rectangle packing problem to generate a schedule for each worker QPU. Afterwards, the Execution Handler then combines the circuits according to the schedule into a new QASM file, creating a new classical register for each circuit and inserting qubit reset operations after every measurement. Finally, the handler sends the QASM files to the QPU interface for execution and unbundles the results before it can be reported back to the users.  In this work, we assess the fidelity of our combined circuits using Quantinuum's H2-Emulator, \cite{quantinuum_h2_emu} as well as the MTS-QCCD compiler and QCCDSim trapped-ion simulator \cite{saki2022muzzle, murali2020architecting}.

\subsection{Packing Formulation}


\begin{table}[h]
\caption{Notation Table}
    \centering
    \begin{tabular}{|c|p{6cm}|}
        \hline
        \textbf{Parameter} & \textbf{Description} \\ \hline
        $N$              & Total number of circuits to be scheduled. \\ \hline
        $W$              & Total number of quantum workers available. \\ \hline
        $Q_w$            & Number of qubits available to quantum worker $w \in \{1, \dots, W\}$. \\ \hline
        $Q_{\text{total}}$ & Total number of qubits in the system, i.e., $\sum_{w=1}^W Q_w$. \\ \hline
        $q_i$            & Number of qubits required by quantum circuit $i$, representing the height of the rectangle. \\ \hline
        $d_i$            & Depth (number of layers) required by quantum circuit $i$, representing the width of the rectangle. \\ \hline
        $C_i$            & Number of 2-qubit gates in quantum circuit $i$ \\ \hline
        $s_i$            & Start time of quantum circuit $i$. \\ \hline
        $e_i$            & End time of quantum circuit $i$, where $e_i = s_i + d_i$. \\ \hline
        $T_w$            & Total execution time (depth) of worker $w$. \\ \hline
        $x_{ij} \in \{0, 1\}$ & Binary variable indicating if quantum circuit $i$ is assigned to physical qubit $j$. \\ \hline
        $z_{iw} \in \{0, 1\}$ & Binary variable indicating if quantum circuit $i$ is assigned to worker $w$. \\ \hline
        $q_{\text{active}}(t)$ & Number of qubits actively involved in non-idling gates at time $t$. \\ \hline
    \end{tabular}
    
    \label{tab:notation}
\end{table}



We now formally define the problem of quantum circuit scheduling, taking inspiration from rectangle packing, and applying this towards scheduling on multiple quantum workers. We treat quantum circuits as rectangles that need to be packed efficiently into a 2D qubit-time grid:

\begin{itemize}
    \item \textbf{Rectangle Height}: The length of the rectangle side along the y-axis, referred to as its height, corresponds to the circuit's width.

    \item \textbf{Rectangle Width}: The side along the x-axis, referred to as the rectangle width, corresponds to the circuit's depth. This is not to be confused with the circuit width.
        
    \item \textbf{Container}: The container for this packing problem is the qubit-time grid, where the container height is equal to the total number of qubits available, and the container width represents the total execution depth (time).
\end{itemize}

\subsection{Objective Functions}
\subsubsection{Maximizing Qubit Utilization}
\begin{equation}
    \max \sum_{w=1}^{W} \sum_{i=1}^{N} \sum_{j=1}^{Q_w} x_{ij} q_i z_{iw}
    \label{eq:max_utilization}
\end{equation}

This equation maximizes the total number of active qubits across all workers. For each worker $w$, it sums up the qubit requirements ($q_i$) of assigned circuits ($z_{iw}=1$) that are mapped to specific qubits ($x_{ij}=1$).

\subsubsection{Minimizing Total Packing Width}
\begin{equation}
    \min T_{\text{max}} = \min \left( \max_{w=1}^W \left( \max_{i=1}^N (e_i \cdot z_{iw}) \right) \right)
    \label{eq:min_depth}
\end{equation}

This objective minimizes the maximum completion time across all workers, also known as the \textit{makespan}. It finds the latest end time ($e_i$) among all circuits assigned to each worker ($z_{iw}=1$) and minimizes the maximum of these times.

\subsection{Qubit Utilization Metrics}

The instantaneous qubit utilization at time $t$ is given by:
\begin{equation}
    U(t) = \frac{q_{\text{active}}(t)}{Q_{\text{total}}}
    \label{eq:utilization}
\end{equation}

This metric represents the fraction of qubits that are actively performing computations (non-idle) at any given time $t$, normalized by the total number of available qubits.

The average utilization over time is calculated as:
\begin{equation}
    U_{\text{avg}} = \frac{1}{T} \sum_{t=0}^{T} U(t)
    \label{eq:avg_utilization}
\end{equation}

This equation computes the time-averaged qubit utilization across the entire execution period, providing a single metric for overall worker efficiency. It can also be shown that this is equivalent to packing efficiency in the Gantt chart representation which is the sum of all circuit areas divided by the container area.

\subsection{Constraints}

\subsubsection{Non-Overlap Constraint}
\begin{equation}
    \sum_{i=1}^{N} x_{ij} \leq 1 \quad \forall j \in \{1, \dots, Q_w\}
    \label{eq:non_overlap}
\end{equation}

This constraint prevents resource conflicts by ensuring each qubit $j$ is assigned to at most one quantum circuit at any time.

\subsubsection{Qubit Capacity Constraint}
\begin{equation}
    \sum_{i=1}^{N} z_{iw} q_i \leq Q_w \quad \forall w \in \{1, \dots, W\}
    \label{eq:capacity}
\end{equation}

This constraint ensures that the total number of qubits required by all circuits assigned to worker $w$ does not exceed its qubit capacity.



\subsubsection{2-Qubit Gate Cutoff Constraint}
\begin{equation}
    \sum_{w=1}^{W} \sum_{i=1}^{N} \sum_{j=1}^{Q_w} C_i x_{ij} z_{iw} \leq \alpha
    \label{eq:cnot_cutoff}
\end{equation}

This constraint ensures that the number of 2-qubit gates in the packed circuit does not exceed the cutoff value to mitigate fidelity loss to ion shuttling.





\section{Algorithm Design}


Our system employs a two-phased, hierarchical scheduling approach to optimize circuit scheduling across multiple quantum workers. Phase-1, a system-wide optimization algorithm (Algorithm~\ref{alg:sys}) determines a balanced assignment of quantum circuits to workers. Then, at Phase-2, the worker-level algorithm (Algorithm~\ref{alg:worker}) handles the detailed scheduling of quantum gates using a rectangle packing approach.

\subsection{System-wide Assignment}

At phase-1, we assign approximately equal subsets of circuits to each quantum worker. This decomposition offers two main benefits: First, each worker’s schedule becomes an independent subproblem which can be solved in parallel on classical resources. 
Second, it enables rapid schedule generation by relaxing global optimality constraints. Evaluating each circuit against all workers to identify the “best” placement increases the scheduling overhead which scales on the number of workers available. Our top-down strategy follows a common practice in distributed and cloud systems, where hierarchical schedulers are used to achieve load balancing and fair resource allocation across workers.

The input consists of a list of quantum circuits \( C = \{c_1, c_2, \dots, c_N\} \), where each circuit \( c_i \) is characterized by \( q_i \) (the number of qubits required) and \( d_i \) (the depth, or number of layers required for the circuit's execution). Additionally, the input includes a list of quantum workers \( W = \{w_1, w_2, \dots, w_W\} \), where each worker \( w_j \) has \( Q_w \).
Algorithm~\ref{alg:sys} uses circuit area, defined as $a_i = d_i \times q_i$, as the metric for load balancing. During operation, the algorithm greedily assigns the current circuit to the worker with the lowest total circuit area. The output of this algorithm are worker-level circuit assignments.

\begin{algorithm}
\caption{Balanced System-Wide Circuit Assignment}
\begin{algorithmic}[1]
\State \textbf{Input:} Circuits \( C = \{c_1, c_2, \dots, c_N\} \), Workers \( W = \{w_1, w_2, \dots, w_W\} \)
\State \textbf{Output:} System-wide assignment matrix \( z_{iw} \)
\State Initialize \( z_{iw} \leftarrow 0 \) for all \( i, w \)
\State Initialize \( L_{w} \leftarrow 0 \) for all \(w \) \Comment{Running total of load per worker}

\For{each circuit \( c_i \in C \)}
    \State \( q_i, d_i \leftarrow \text{qubit count, depth of circuit } c_i \)
    \State \( a_i \leftarrow q_i \times d_i\)
    \State \(\text{min\_load} \leftarrow min(L_j) \forall j \in W\)
    
    \For{each worker \( w_j \in W \)}
        \If{\( L_j = \text{min\_load}\)} 
            \State \( L_j \leftarrow L_j + a_i\) \Comment{Update worker load total}
            \State \( z_{i,j} \leftarrow 1 \) \Comment{Update circuit-worker assignment}
            \State \textbf{break}
        \EndIf
    \EndFor    
\EndFor
\For{each worker \( w_j \in W \)}
    \State Call Algorithm~\ref{alg:worker} on \( w_j\) using  \(z_{iw} \)
\EndFor
\State \Return \(z_{iw} \)
\end{algorithmic}
\label{alg:sys}
\end{algorithm}

\subsection{Worker-Level Circuit Scheduling}

Algorithm~\ref{alg:worker} is responsible for generating a circuit schedule of a quantum worker given the assignments from Algorithm \ref{alg:sys}. Given that the qubit capacity and circuit set are fixed, minimizing the makespan (Equation \eqref{eq:min_depth}) has the effect of maximizing qubit utilization (Equation \eqref{eq:avg_utilization}). A modified Skyline heuristic algorithm for rectangle packing \cite{jylanki2010thousand} is used to determine the exact circuit placements on the qubit-time grid while following the non-overlap and qubit capacity constraints (Equations \eqref{eq:non_overlap}, \eqref{eq:capacity}).

The input consists of the list of circuits assigned to this worker,~ \( C = \{c_1, c_2, \dots, c_N\} \), where each circuit \( c_i \) is characterized by \( q_i \) (the number of qubits required) and \( d_i \) (the depth of the circuit). A worker \( w_j \) has available qubits \( Q_w \) modeled as \( \{Tr_1,Tr_2,...,Tr_k\}\), specifying the capacities on the trap level.
The output is a detailed schedule that lists the qubit assignment and schedule depth for each circuit, which can be thought of as the coordinates in the qubit-time grid. Being a heuristic-based method, Algorithm 2 places a local priority on minimizing the packing width (Equation \eqref{eq:min_depth}) for each circuit and we evaluate the result by measuring qubit utilization (Equation \eqref{eq:max_utilization}) averaged over all layers.

\begin{algorithm}
\caption{Maximize Utilization for Worker \( w \)}
\begin{algorithmic}[1]
\State \textbf{Input:} Circuits \( C = \{c_1, c_2, \dots, c_N\} \), Worker \( w_j \) with \( Q_w \) qubits and maximum depth \( T_{\text{max}} \)
\State \textbf{Output:} Schedule of gates, assignment matrices \( x_{ij} \), and resource utilization \( U_w \)
\State \textbf{Initialize:} Sort circuits \( C \) by qubit requirement \( q_i \) in descending order
\State Initialize output list and skyline grid \( Output, G[Tr_k][Q_w] \leftarrow \text{False} \)
\State Initialize assignment matrix \( x_{ij} \leftarrow 0 \) for all \( i, j \)
\State Initialize \( U_w \leftarrow 0 \), \( q_{\text{active}}(t) \leftarrow 0 \) for all \( t \)
\State Initialize \( T_w \leftarrow 0 \) \Comment{Track worker's execution time}
\State Initialize \( CNOT \leftarrow 0 \) \Comment{Current 2-qubit gate running total}

\For{each circuit \( c_i \in C \)}
    \State \( q_i, d_i, C_i \leftarrow \text{qubit count, depth, 2Q gate count of circuit } c_i \)
    \State \( min_i \leftarrow 0 \) \Comment{Keep track of minimum layer  for circuit placement}
    \State \( Tr_i, qs_i \leftarrow 0 \) \Comment{Trap number and qubit start number}
    \For{each Trap \( Tr_k \in Tr \)} \Comment{Iterate on each trap}
        \State \(min_{cur}, qs_{cur} \leftarrow minimax(G[Tr_k],q_i)\) \Comment{Check packing width for current trap assignment while considering the constraints, Equations (\ref{eq:min_depth}, \ref{eq:non_overlap}, \ref{eq:capacity})}
        \If{\( min_{cur} < min_i\)}
            \State \(min_i = min_{cur}\) \Comment{Update packing width}
            \State \(Tr_i = Tr_k\) \Comment{Update trap number}
            \State \(qs_i = qs_{cur}\) \Comment{Update qubit start number}
        \EndIf
    \EndFor
    
    \State $G[Tr_i][min_i: min_i+q_i] = min_i + d_i$ \Comment Update skyline grid with new depth
    \State $U_W += a_i$ \Comment{Update utilization Equation \eqref{eq:avg_utilization}}
    \State $CNOT += C_i$ \Comment{Update running 2Q gate count}
    \State $Output += (i,Tr_i,qs_i, min_i)$  \Comment{Record the trap number, qubit start, and depth start for $c_i$}
    \State $x_{iw} \leftarrow 1$ \Comment{Update circuit assignment for Worker $w$}
    
    \If {$CNOT \geq \alpha$} \Comment{Check 2Q constraint Equation \eqref{eq:cnot_cutoff}}
        \State Record cutoff in $Output$
        \State Reset $G$, $CNOT$
    \EndIf
\EndFor

\State $ T_w = max(G)$
\State Calculate final \( U_w  = \frac{U_w}{T_w \times q_i}\) using Equation (\ref{eq:avg_utilization})
\State \Return \( Output, U_w \)
\end{algorithmic}
\label{alg:worker}
\end{algorithm}

The algorithm begins by sorting the circuit list input in descending width order, initializing the running count of 2-qubit gates for the cutoff, and maintains a "skyline" list for each trap, recording the current schedule depth on the individual qubit level. A list structure is also prepared as the container for the output.

The main packing logic consists of checking the best circuit placement across each trap that minimizes the schedule depth while keeping track of the 2-qubit gate count. When Equation \eqref{eq:cnot_cutoff} is met, a new subschedule is created and the packing restarts from an empty container.

When all circuits have been procesed, the final output is the collection of subschedules and the average utilization of the combined schedule.

\section{Evaluation}
We assess the effectiveness of our framework to improve quantum cloud operations in four key areas: (1) single-worker fidelity, (2) packing efficiency, (3) shuttling operations, and (4) multi-worker task allocation. 
In each scenario, we modeled a 20-qubit QCCD trapped ion device, to match the capacity of the Quantinuum System Model H2 emulator \cite{quantinuum_h2_emu}. The default values of the noise model's parameters that were used during testing can be found in Table \ref{tab:h2_noise_params}.

\begin{table}[h]
\caption{Quantinuum H2-Emulator default noise model parameters}
    \centering
    \begin{tabular}{|c|c|}
        \hline
        \textbf{Parameter} & \textbf{Value}\\ \hline
        p1 & 7.3e-05\\ \hline
        p2 & 0.00129\\ \hline
        p\_meas (0 outcome) & 0.0009\\ \hline
        p\_meas (1 outcome) & 0.0018\\ \hline
        p\_crosstalk\_meas & 8.8e-06\\ \hline
        p\_init & 4e-05\\ \hline
        p\_crosstalk\_init & 9.6e-06\\ \hline
        p1\_emission\_ratio & 0.32\\ \hline
        p2\_emission\_ratio & 0.59\\ \hline
        quadratic\_dephasing\_rate & 0.043\\ \hline
        linear\_dephasing\_rate & 0.0028\\ \hline
        coherent\_to\_incoherent\_factor & 2.0\\ \hline
        przz\_a & 1.518\\ \hline
        przz\_b & 0.241\\ \hline
        przz\_c & 1.518\\ \hline
        przz\_d & 0.241\\ \hline
        przz\_power & 1.0\\ \hline
    \end{tabular}
    \label{tab:h2_noise_params}
\end{table}


\subsection{Metrics}
Previous QMP studies put more emphasis on measuring the fidelity of the result than quantifying improvements to hardware resource utilization. Given this gap, we define new evaluation metrics that capture both spatial and temporal aspects of qubit operation.

\begin{enumerate}
    \item \textbf{Fidelity}:
        We use the fidelity result from the error model of QCCDSim \cite{murali2020architecting} based on the number of shuttling and gate operations performed \cite{murali2020architecting}, and the
        the Probability of Successful Trial (PST) result using the H2-Emulator \cite{quantinuum_h2_emu} as a shot-based estimator of circuit fidelity, as used in previous QMP studies \cite{das2019case, niu2023enabling, ohkura2022simultaneous, liu2024qucloud+}.
        \begin{equation}
            PST = \frac{\text{number of successful trials}}{\text{total number of trials}} \times 100\%
            \label{eq:pst}
        \end{equation}
        
    \item \textbf{Shuttling count}:
        The is the number of shuttles needed to execute a circuit schedule and is dependent on the mapping of circuit qubits to the trap topology, obtained using the MTS-QCCD compiler \cite{saki2022muzzle}. Because QCCD devices accumulate heat and motion during each shuttle, the reliability of future gate operations are degraded. Therefore, these must be kept at a minimum.


    \item \textbf{Average Utilization}:
        As defined in Equation (\ref{eq:avg_utilization}), this metric quantifies the complete utilization of a circuit schedule as the total count of active qubit gates over the total space-time qubit grid area used. An average utilization value of 80\% would mean that 80\% of the qubit-time grid is composed of gate operations and 20\% idle gates.
    
    \item \textbf{Layer Reduction Factor (LRF)}:
        This quantifies the reduction of the number of layers due to multi-programming as opposed to serial execution. It is calculated by the ratio of the schedule makespan to the total schedule depth if the circuits were run in series.
        \begin{equation}
            LRF = \frac{L_{serial}-L_{makespan}}{L_{serial}} \times 100\%
            \label{eq:lrf}
        \end{equation}
        It is modeled similar to the Trial Reduction Factor metric that quantifies the reduction in shot count due to QMP. \cite{das2019case}


    \item \textbf{Scheduling Overhead}:
        This is the processing time taken by the classical computer to compute the packing layout.
        
        
\end{enumerate}

\subsection{Benchmarks}

These benchmark circuits, sourced from Revlib \cite{revlib} and QASMBench \cite{qasmbench}, are chosen to allow for comparison of our results with previous QMP studies, and to fit within the limited capacity of the H2-Emulator.
These circuits also produce only a single outcome in noiseless simulations, which simplifies the PST calculations.
Table \ref{tab:benchmark_circuits} contains the details of each circuit, and we classify circuits numbered 1-8 as small circuits while numbers 9-14 are classified as medium-sized circuits.



\begin{table}[h]
\caption{Benchmark circuit details.}
\centering

\begin{tabular}{|r|l|r|r|r|r|}
\hline
\textbf{ID} & \textbf{Circuit} & \textbf{Width} & \textbf{Depth} & \textbf{2Q Gates} \\
\hline
1 & grover\_n2      & 2  & 16 & 2\\
\hline
2 & fredkin\_n3     & 3 & 19 & 8\\
\hline
3 & toffoli\_n3     & 3 & 18 & 6\\
\hline
4 & 3\_17\_13       & 3 & 23 & 17\\
\hline
5 & adder\_n4       & 4 & 23 & 10\\
\hline
6 & 4mod5-v1\_22    & 5 & 13 & 11\\
\hline
7 & 4gt13\_92       & 5 & 39 & 30\\
\hline
8 & mod5mils\_65    & 5 & 22 & 16\\ 
\hline
9  & adder\_n10           & 10 & 142 & 65\\
\hline
10 & multiply\_n13        & 13 & 98 & 40\\
\hline
11 & bv\_n14              & 14 & 41 & 13\\
\hline
12 & multiplier\_n15      & 15 & 574 & 246\\
\hline
13 & qec9xz\_n17          & 17 & 53 & 32\\
\hline
14 & bigadder\_n18        & 18 & 284 & 130\\
\hline                         
\end{tabular}
\label{tab:benchmark_circuits}
\end{table}

\subsection{Results}
To evaluate the efficiency of our custom packing algorithm, we introduce other simple packing algorithms to establish points of comparison.

\begin{itemize}
    \item \textbf{Serial}-- this is the baseline "packing" obtained by scheduling the circuits one at a time. It models the single-tenant execution pattern of current quantum cloud services.
    \item \textbf{First-in First-out (FIFO)}-- this is a simple greedy packing algorithm that packs each incoming circuit in the earliest possible layer.
    \item \textbf{RectPack} -- this is the packing result obtained from the \textit{rectpack} Python library\cite{rectpack}. It implements the heuristic algorithms described in a survey paper \cite{jylanki2010thousand}.
    \item \textbf{CircPack}-- this is our algorithm implementation described in Section 5, written in Python 3.12.9 using pytket v. 1.33.1.

    \item \textbf{QuMC~\cite{niu2023enabling}} -- a superconducting-based compiler that partitions qubits into crosstalk-aware regions for QMP
    
    \item \textbf{QuCloud+~\cite{liu2024qucloud+}}-- a superconducting-based and fidelity-driven solution.

    \item  \textbf{Quantinuum 2D Packing~\cite{quantinuum_circuit_stitching}} -- Quantinuum's depth-aligned solution for QMP based on their "Circuit Stitching" knowledge article.
    
\end{itemize}

The results obtained from our packing alogrithms, evaluations using MTS-QCCD and QCCDSim, and simulated multi-QPU scheduling were executed using a device with an AMD Ryzen 7 7840HS at 3.80 GHz and 32 GB of RAM.

\begin{table*}
\caption{Comparison of different packing algorithms. The fidelity values were obtained using the QCCDSim \cite{murali2020architecting}.}
\centering
\begin{tabular}{cl|cccc|ccc}
\toprule
\textbf{Size} & \textbf{Algorithm} & \textbf{Makespan} & \textbf{No. of Cutoffs} & \textbf{Shuttles} & \textbf{Fidelity} & \textbf{Avg. Util.} & \textbf{LRF} & \textbf{Time} \\
\midrule

\multirow{4}{*}{20}     & Serial    & 395       & 20      & 0     & 99.72\%   & 21.60\%   & ---      & N/A \\
                        & FIFO      & 120       & 0     & 2     & 94.32\%   & 71.12\%   & 69.62\%  & \textbf{0.0384 s} \\
                        & RectPack  & \textbf{91}        & 0      & 7     & 93.63\%   &\textbf{ 93.79\%}   & \textbf{76.96\%}  & 0.0405 s \\
                        & \textbf{CircPack}  & 97        & 0      & \textbf{0}     & \textbf{94.50\%}   & 87.99\%   & 75.44\%  & 0.1989 s \\
\midrule

\multirow{4}{*}{100}    & Serial    & 1757      & 100     & 0     & 99.75\%   & 20.46\%   & ---      & N/A \\
                        & FIFO      & 422       & 0     & 30    & 71.73\%   & 85.18\%   & 75.98\%  & 0.3657 s \\
                        & RectPack  & \textbf{379}       & 0     & 23    & 74.04\%   & \textbf{94.84\%}   & \textbf{78.43\%}  & \textbf{0.2133 s} \\
                        & \textbf{CircPack}  & 405       & \textbf{4}     & \textbf{1}     & \textbf{93.92\%}   & 88.75\%   & 76.95\%  & 0.5052 s \\
\midrule

\multirow{4}{*}{150}    & Serial    & 2620      & 150      & 0     & 99.76\%   & 19.50\%   & ---      & N/A \\
                        & FIFO      & 596       & 0     & 33    & 63.98\%   & 85.72\%   & 77.25\%  & 0.3749 s \\
                        & RectPack  & \textbf{529}       & 0     & 19    & 66.76\%   & \textbf{96.58\%}   & \textbf{79.81\%}  & \textbf{0.2820 s} \\
                        & \textbf{CircPack}  & 575       & \textbf{5}     & \textbf{3}     & \textbf{93.16\%}   & 88.85\%   & 78.05\%  & 0.7404 s \\
\midrule

\multirow{4}{*}{200}    & Serial    & 3762      & 200      & 0     & 99.74\%   & 20.24\%   & ---      & N/A \\
                        & FIFO      & 870       & 0     & 41    & 49.38\%   & 87.51\%   & 77.41\%  & \textbf{0.5149 s} \\
                        & RectPack  & \textbf{779}       & 0     & 20    & 54.84\%   & \textbf{97.73\%}   & \textbf{79.29\%}  & 0.5643 s \\
                        & \textbf{CircPack}  & 852       & \textbf{8}     & \textbf{2}     & \textbf{93.66\%}   & 89.35\%   & 77.35\%  & 0.8906 s \\
\bottomrule
\end{tabular}
\label{table: pack_efficiency}
\end{table*}

\subsubsection{Single Worker Packing Comparisons}

To simulate a quantum cloud job queue, we randomly sampled small circuits from our benchmark set and evaluated four queue sizes: 20, 100, 150, and 200 circuits, testing performance under varying workload conditions. Using a simulated linear two-trap layout with 10 ions per trap, we applied the Serial, FIFO, RectPack, and CircPack methods to generate circuit schedules. Each schedule was evaluated in terms of makespan, number of cutoffs, shuttle count, and fidelity, along with the average utilization, layer reduction factor (LRF), and scheduling time.

The complete results are presented in Table \ref{table: pack_efficiency}, with the best-performing value in each column (excluding the Serial baseline) highlighted. Obviously, Serial schedules achieve the best fidelity across all sizes at around 99.70\%, due to minimal shuttles and frequent cutoffs. However, it simply executes circuits one at a time, significantly sacrificing makespan and average resource utilization, e.g., around 20\%. For FIFO and RectPack, the naive packing approach improves the makespan and average utilization, but their fidelity result decreases as the number of circuits in the queue increases. A drop from 94.32\% and 93.63\% for 20 circuits down to 49.38\% and 54.84\% for 200 circuits can be observed in in Table \ref{table: pack_efficiency}. This significant fidelity decrease makes them impractical.

On the other hand, CircPack demonstrates the best trade-off between resource utilization and fidelity. For 200 circuits, it achieves a fidelity of 93.66\%, stemming from its trap-aware design and dynamic cutoffs. These minimize the required shuttling operations and ion transport while making slight reductions in packing performance. Despite the additional constraints, CircPack completes scheduling with an average utilization of 89.35\% and LRF of 77.35\% in only 0.8906 s, an acceptable overhead given its notable fidelity gains. Moreover, for larger workloads, CircPack can be further accelerated through parallelization across multiple quantum workers, where each sub-schedule can be solved in parallel.

Detailed Analysis: Reducing the number of incurred shuttles is key to limiting error accumulation on QCCD trapped ion devices \cite{murali2020architecting, moses2023race}, as the reliability of future gate operations depend on the amount of heat and motion currently present in the ion chain. Therefore, allowing the device to reset between consecutive executions via a schedule cutoff is crucial to avoid accumulating these sources of noise.

The Serial schedule shows how to maximize circuit fidelity performing the minimum number of shuttles and applying a cutoff after each circuit, as observed with a consistent 0 shuttle count and a cutoff values equal to the queue size: 20, 100, 150, 200. The drawback is that it fails to leverage any unused qubit resources.
CircPack approaches the task by implementing the techniques described in Subsections ~\ref{section:trap_based_packing} and~\ref{section:2qcutoff}.
The trap-based packing method minimizes additional shuttles required due to concurrent circuit executions by restricting circuits to be scheduled across traps, as seen in Figure \ref{fig:qccd architecture}(d). As such, shuttle counts remain at 0, 1, 3, and 2, for the different queue sizes. On the other hand, the dynamic schedule cutoff mechanism partitions the schedule before the noise exceeds threshold $\alpha$ as defined in Section \ref{section:2qcutoff}, obtaining cutoff numbers of 0, 4, 5, and 8.  By applying both techniques, the two competing objectives of obtaining reliable results and maximizing hardware capacity are balanced.

In contrast, both FIFO and RectPack focus purely on increasing circuit throughput, e.g., average resource utilization, at the expense of decreasing fidelity with the highest number of shuttles and zero cutoffs. At a queue size of 200 circuits, FIFO reports a makespan of 870 but fidelity result of only 49.38\%. In contrast, RectPack achieves the best makespan result of 779, but a similar dive in fidelity is observed at 54.84\%. These results underscore the importance of reducing shuttle counts and introducing schedule cutoffs to maintaining reliable results. CircPack reduced the number of shuttles from 41, 20 to 2, which represents a 95.12\% and 90.00\% decrease compared to FIFO and RectPack, leading to an increase $44.28\%$ and $32.82\%$ in respective fidelities. CircPack's added cutoffs also contributed this fidelity gain while incurring a minimal decrease of $8.38\%$ in average utilization.

A key restriction to consider is that FIFO’s naive and greedy scheduling does not support cutoff-based partitioning, while RectPack’s opaque interface is inherently incompatible or will require heavy modifications on the source-code level. Introducing the cutoff constraints externally—such as by tracking 2-qubit gate counts before or after packing would distort RectPack’s internal sorting and bin assignment, resulting in an entirely different schedule. In particular, post-hoc subdivision of a completed RectPack schedule would invalidate the original packing result order and require recomputation, yielding either suboptimal results or inconsistent schedules that do not satisfy the non-overlap (Equation \ref{eq:non_overlap}) and qubit capacity (Equation \ref{eq:capacity}) packing constraints.

Overhead Analysis:Scheduling overhead, measured as CPU processing time, scales linearly with queue size across all methods. Since Serial requires no packing, overhead is not applicable. FIFO is the fastest due to its greedy queueing heuristic, followed closely by RectPack, which benefits from a highly-optimized MaxRects algorithm implementation. CircPack incurs the largest overhead because it enforces additional hardware-capacity and trap-based placement constraints, yet even in the worst cases, the gap relative to RectPack remains only 0.4584 s (Size = 150) and 0.3263 s (Size = 200).

\subsubsection{H2-Emulator Fidelity Analysis}

\begin{figure}
    \centering
    \subfigure[One Pack: 20 small circuits]{\includegraphics[width=0.8\linewidth]{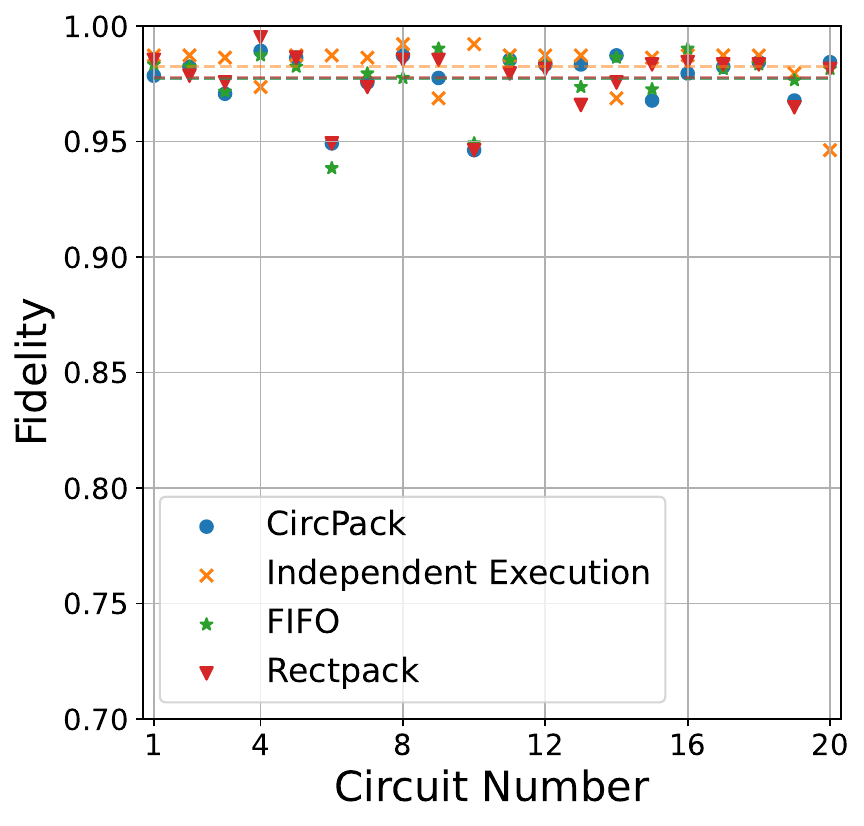}} 
    \subfigure[One Pack: 14 small/medium circuits]{\includegraphics[width=0.8\linewidth]{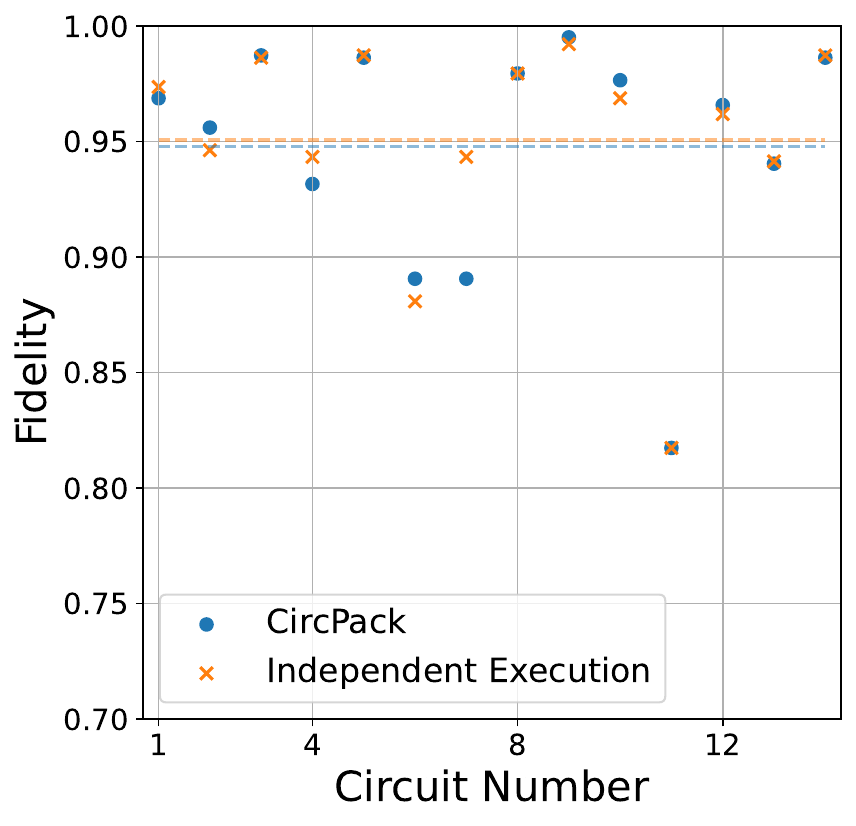}} 
`    \caption{Single-worker circuit reliability measurements on the Quantinuum H2-Emulator.}
\vspace{-0.2in}
    \label{fig:result_pst}
\end{figure}

A fidelity evaluation was conducted using the Quantinuum H2-Emulator \cite{quantinuum_h2_emu}. We first established baseline probability of successful trial (PST) values for each circuit in the benchmark set by executing them independently with 1024 shots each. These baseline PSTs were then compared against those obtained from multi-programmed executions to quantify the fidelity impact of concurrent scheduling.

Two experimental scenarios were tested to simulate job queues of varying circuit diversity: (1) a queue of 20 small circuits, and
(2) a mixed queue of 14 small and medium circuits.
Results are shown as scatter plots in Figure \ref{fig:result_pst} (a) and (b), where each point on the x-axis corresponds to an individual circuit and the y-axis indicates its measured PST. Different algorithms are represented by distinct markers and colors, while dashed lines indicate the average fidelities for each method.

The emulator results indicate that CircPack introduces only a minimal fidelity penalty, consistent with the previous findings by Niu et al. on trapped-ion-based multi-programming \cite{niu2022multi}.
For both queue configurations, CircPack produced PST values that closely matched the results from independent circuit execution: an average fidelity difference of 0.52\% is observed in the first scenario and 0.27\% in the second.

On the other hand, the results do not indicate any significant differences in fidelity across all packing methods at this scale.
As shown in Figure \ref{fig:result_pst} (a), the average fidelities of FIFO, RectPack, and CircPack differ by at most 0.06\%,
Further increasing the number of shots used should improve the accuracy of the simulations and better evaluate the scalability of the methods, however the emulator has a 30-minute classical runtime limit which restricts the maximum shot count and circuit queue size for our tests. This is the cause of the incomplete data for FIFO and RectPack in Figure \ref{fig:result_pst} (b). In contrast, CircPack successfully executed the entire workload by leveraging its 2-qubit gate cutoff, which partitions the schedule into smaller batches that fits within emulator constraints.

Naturally, independent executions, e.g., serial schedules, are expected to yield the most reliable results. The goal of QMP methods, e.g., CircPack, QuMC and QuCloud+, on the other hand, is to minimize the fidelity gap between independent and parallel executions while maintaining substantial throughput improvements. However, some QMP results have large fidelity variations relative to their baselines. In general, added noise in multi-programmed schedules can be attributed to the following reasons:
\begin{enumerate}
    \item Crosstalk noise due to co-occurring 1-qubit and 2-qubit gate operations.
    \item Measurement noise from other circuits during parallel execution.
    \item Additional shuttle operations that were introduced due to the co-execution of multiple circuits.
    \item Accumulated heat and motion in trapped ion chains from shuttling and gate operations that degrade fidelity.
\end{enumerate}

While trapped-ion hardware is regarded for its low crosstalk and high fidelity mid-circuit measurement and reset which reduces the noise classified in (1) and (2), CircPack addresses noise sources (3) and (4) through the use of trap-based packing and the 2-qubit gate scheduling cutoff methods.

\subsubsection{Shuttling Count Trend}

\begin{figure}
    \centering
    \includegraphics[width=0.9\linewidth]{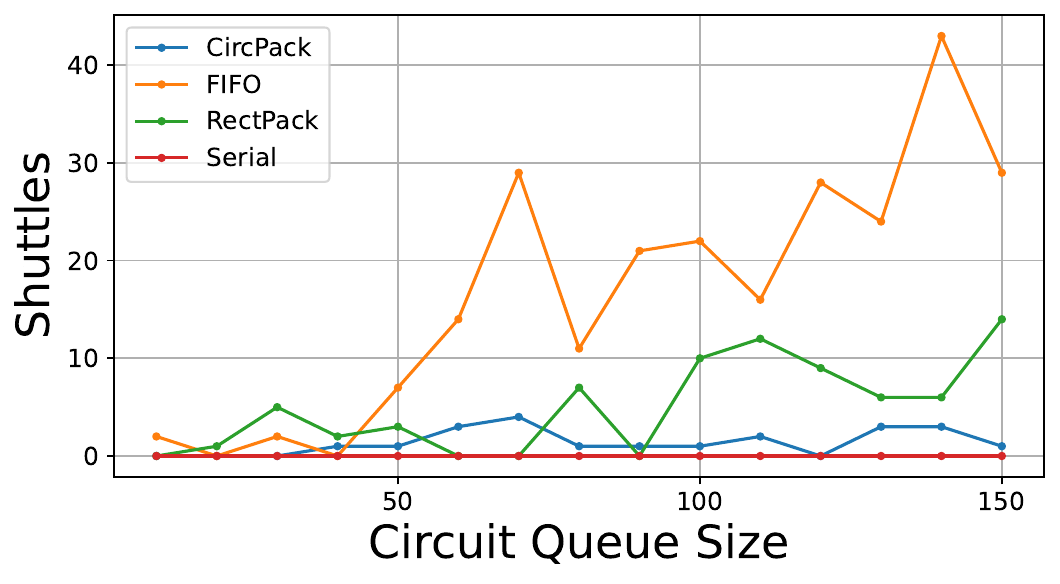}
    \caption{Comparison of the shuttling operation count result using different packing algorithms on the MTS-QCCD compiler\cite{saki2022muzzle}.}
    \label{fig:result_fidelity_degradation}
\end{figure}

To further analyze how shuttling overhead scales with increasing circuit queue sizes, we used the MTS-QCCD compiler \cite{saki2022muzzle} to measure shuttle counts under varying workload conditions. Assuming a linear two-trap layout with 10 ions per trap, we simulated queue sizes ranging from 10 to 150 randomly sampled small circuits from our benchmark set. The resulting shuttling trends are presented in Figure \ref{fig:result_fidelity_degradation}.


With the exception of the Serial baseline, which has no parallelism and has no shuttling as the small circuits all fit in a single trap, all methods show a general upward trend in shuttle counts as the number of circuits increases. The results also display moderate variance between adjacent queue sizes, reflecting the dependence of shuttling behavior on the specific spatial distribution and qubit connectivity of the circuits within each queue.

Among the evaluated methods, CircPack consistently achieves the fewest shuttling operations across all queue sizes. At the largest tested workload of 150 circuits, CircPack incurred only a single shuttle operation, compared to 14 for RectPack and 29 for FIFO, an improvement attributed to CircPack’s trap-aware packing strategy. By reducing the number of circuit layers that span across traps, CircPack effectively limits the need for ion transport and thus maintains higher overall reliability.

\subsubsection{Comparison with Previous Work}
Ideally, throughput benchmarks for quantum multiprogramming (QMP) methods would be based on actual runtime measurements on hardware. However, prior studies did not report any runtime data, and the H2-Emulator does not provide runtime estimates for real devices. Moreover, execution duration on QCCD architectures cannot be inferred from schedule makespan and gate counts alone, since the shuttling operation counts and total distance, which are heavily dependent on the device configuration and compiler used, dominate total runtime \cite{moses2023race}. Since information on the H2-emulator's shuttling compiler is not publicly available, we compare methods primarily from a scheduling perspective.

Among superconducting (SC)-based approaches, we use QuCloud+ \cite{liu2024qucloud+} and QuMC \cite{niu2023enabling} to serve as our baselines for comparison as both circuit workload and fidelity results are available. Palloq \cite{ohkura2022simultaneous}, on the other hand, failed to include exact counts per circuit type used in its evaluation. For trapped-ion (trapped-ion)-based methods, we benchmark CircPack against the circuit-stitching approach (here referred to as Quantinuum 2D Packing), using a set of 24 equal-width small circuits.

Table \ref{table: qumc_comparison} summarizes results for QuMC, QuCloud+, Quantinuum 2D Packing, and CircPack. We used the circuit mixes and fidelity data from the original publications, as replicating the study proved to be difficult due to major version changes in Qiskit. Additionally, the original devices used in the study have already been retired: QuCloud+ used the 27-qubit IBMQ Toronto, and QuMC used the 65-qubit IBMQ Manhattan, as opposed to the 20-qubit H2-Emulator used for Quantinuum 2D Packing and CircPack. To resolve the differences in circuit workloads and device capacities, we perform three separate comparisons against CircPack.

As shown in the results, CircPack consistently outperforms existing QMP algorithms across all comparison scenarios. The largest relative improvements are observed against QuCloud+, with gains of up to $70.72\%$ in fidelity, $62.67\%$ in utilization, and $32.80\%$ in LRF. This substantial performance gap is attributed to the limited parallelism of QuCloud+ to only two circuits, resulting in low utilization scores and the noisy nature of SC qubits which degrade fidelity. Next, QuMC achieves a marginally shorter makespan and a higher LRF value compared to CircPack, due to using a 65-qubit device which allowed more parallel executions. However, this advantage comes at the expense of inefficient qubit–time utilization, with QuMC achieving only $27.28\%$ average utilization versus CircPack’s $64.33\%$.
Finally, Quantinuum 2D Packing slightly exceeds CircPack's fidelity result by a negligible $0.06\%$, which is too small to classify as a practical performance advantage. Instead, this result highlights the high-fidelity nature of trapped-ion platforms, as both algorithms utilized the H2-Emulator and achieved higher than $98\%$ fidelity. Nonetheless, the Quantinuum method exhibits limited generality, as it requires all input circuits to share identical widths, and performs worse in makespan, utilization, and LRF relative to CircPack.

\begin{table}
\centering
\caption{Comparison of different QMP frameworks.}
\begin{tabular}{lrrrr}
\toprule
\textbf{Algorithm}& \textbf{Makespan} & \multicolumn{1}{c}{\textbf{Avg. Util.}} & \textbf{Avg. Fid.} & \multicolumn{1}{c}{\textbf{LRF}} \\
\midrule
QuCloud+ & 108 & 15.05\% &   25.59\% & 42.86\% \\
\textbf{CircPack} & \textbf{46} & \textbf{77.72\%} & \textbf{96.31\%} &\textbf{ 75.66\%} \\
\midrule
QuMC & \textbf{116} & 27.28\% &   $\sim$ 55.5\% &\textbf{ 67.32\% } \\
\textbf{CircPack} & 120 & \textbf{64.33\%} & \textbf{96.91\%} & 66.20\% \\
\midrule
Quantinuum  & \multirow{2}{*}{184} & \multirow{2}{*}{33.59\%} & \multirow{2}{*}{\textbf{ 98.34\%}} & \multirow{2}{*}{55.34\%} \\
2D Packing &  &  &  &  \\
\textbf{CircPack} & \textbf{92} & \textbf{67.17\%} & 98.28\% &\textbf{77.67\%} \\
\bottomrule
\end{tabular}
\label{table: qumc_comparison}
\end{table}

\subsubsection{Multi-worker Resource Assignment}
Assigning jobs to multiple workers requires an additional load balancing constraint, making sure that no single worker has a qubit utilization and makespan that are significantly higher than the rest. To test this, we prepare a simulated queue of 1000, 2000, and 3000 circuits and distribute them to 5 identical 20-qubit worker QPUs. The results, as seen in Table \ref{table: multi_worker}, show well-balanced loads for each of the 5 workers: a 2.83\% decrease in makespan and 1.74\% difference in average utilization between the highest and lowest load workers. 
This demonstrates that an equal resource allocation is possible even for large workloads despite using relatively simple heuristics.

\begin{table}
\centering
\caption{Result of distributed circuit scheduling on different circuit queue sizes.}
\begin{tabular}{crrrrr}
\toprule
\textbf{Size} & \textbf{Worker} & \textbf{Circuits} & \textbf{Makespan} & \multicolumn{1}{c}{\textbf{Avg. Util.}} & \textbf{Time}\\
\midrule
\multirow{ 5}{*}{1000} 
& 1 & 199 & 9303 & 70.94\% &  1.5223 s\\
& 2 & 200 & 9084 & 71.04\% &  1.3240 s\\
& 3 & 176 & 9139 & 70.68\% &  1.6150 s\\
& 4 & 201 & 9471 & 68.04\% &  1.3703 s\\
& 5 & 224 & 9452 & 68.49\% &  1.3421 s\\
\midrule
\multirow{ 5}{*}{2000} 
& 1 & 372 & 18855 & 73.43\% &  2.6907 s\\
& 2 & 409 & 18909 & 73.13\% &  3.1750 s\\
& 3 & 379 & 18957 & 72.83\% &  3.1180 s\\
& 4 & 411 & 19510 & 70.90\% &  2.9235 s\\
& 5 & 429 & 18942 & 72.86\% &  3.2035 s\\
\midrule
\multirow{ 5}{*}{3000} 
& 1 & 645 & 28220 & 70.68\% &  4.9165 s\\
& 2 & 562 & 27420 & 72.42\% &  4.4243 s\\
& 3 & 594 & 28023 & 71.11\% &  4.7676 s\\
& 4 & 614 & 27982 & 70.77\% &  4.9187 s\\
& 5 & 585 & 27703 & 71.81\% &  4.1147 s\\

\bottomrule
\end{tabular}
\label{table: multi_worker}
\end{table}

\section{Conclusion and Discussion}
As quantum computing becomes increasingly accessible through cloud platforms, the challenge of managing scarce hardware resources grows correspondingly. This paper presents CircPack, a hardware-aware circuit packing framework tailored to QCCD trapped-ion architectures, designed to improve throughput in quantum cloud environments. CircPack enables efficient static scheduling while minimizing shuttle operations, achieving high execution fidelity with minimal classical processing overhead.

Relative to RectPack, a generic rectangle-packing solver, CircPack generates hardware-aware schedules with comparable makespans but with fewer shuttle operations, though at the cost of extra CPU time from added constraints. Using both the H2-emulator and QCCDSim, we observe that CircPack achieves a $38.82\%$ increase in fidelity despite the $8.38\%$ decrease in utilization and a $0.3263$ s increase in processing time for a 200-circuit queue. These improvements stem from its trap-based packing strategy and 2-qubit gate cutoff mechanism, which reduce ion transport and accumulated noise.

Compared  to prior quantum multiprogramming (QMP) approaches, our results highlight the advantage of trapped-ion (TI) systems over superconducting (SC) platforms due to their higher qubit reliability and more efficient space–time utilization: up to $70.72\%$ improvement to fidelity, $62.67\%$ increase in average utilization, and $32.80\%$ improvement in total circuit layer reduction. To our knowledge, this is the first study to integrate temporal qubit reuse and evaluate multi-worker load balancing within QMP workflows.


Future work will focus on scaling CircPack to larger circuits and longer queues evaluated on physical QCCD hardware, with empirical execution-time measurements. Additional directions include incorporating security and privacy constraints into QMP frameworks; generalizing the packing model to other all-to-all architectures such as neutral atoms; developing adaptive shuttling reduction strategies for circuits that exceed the trap size; and exploring circuit transformations that reduce depth and/or width for more compact, noise-resilient packings.

\bibliographystyle{IEEEtran}
\bibliography{references}
\vskip -2.5\baselineskip plus -1fil
\begin{IEEEbiography}[{\includegraphics[width=1in,height=1.25in,clip,keepaspectratio]{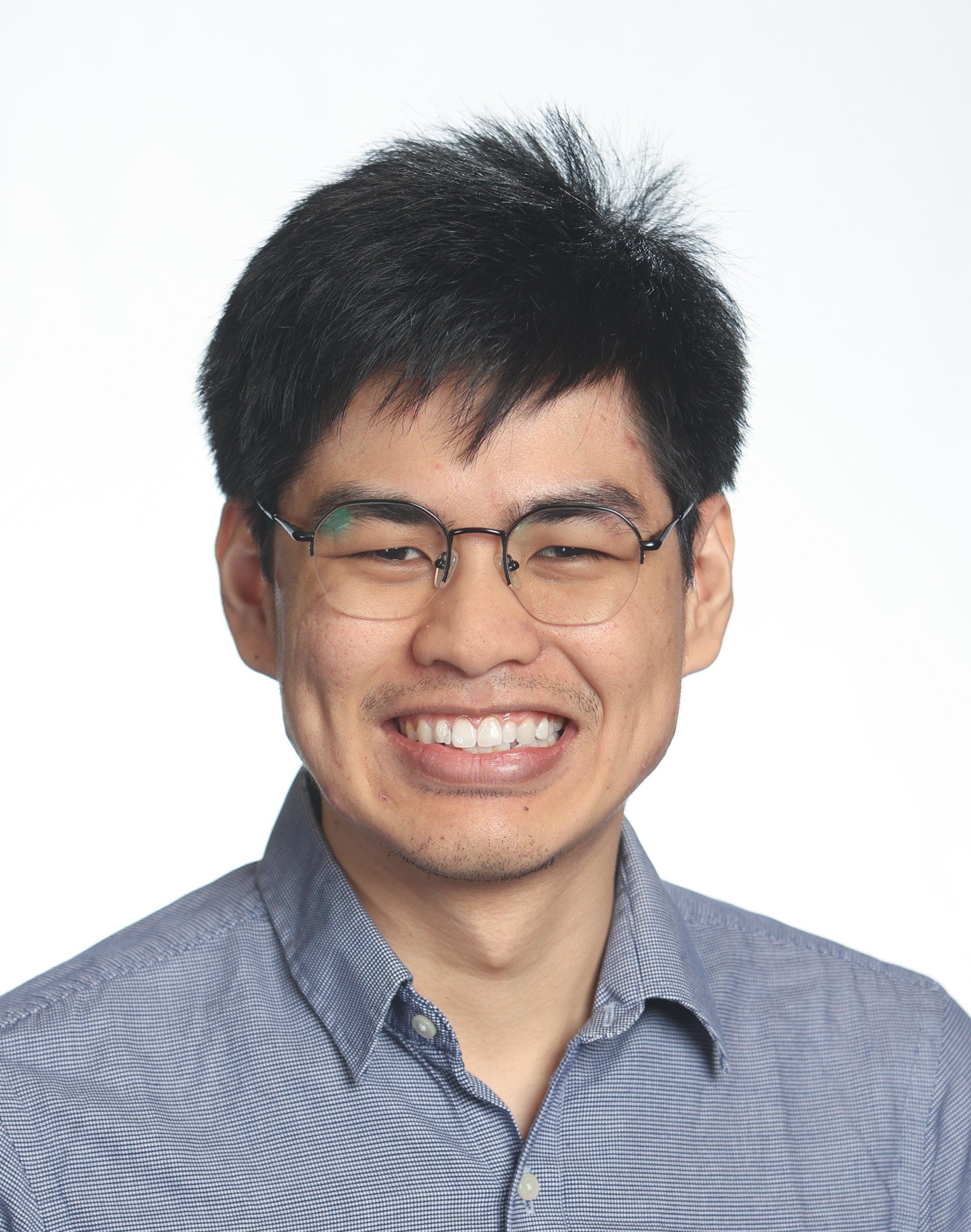}}]%
{\textbf{MIGUEL PALMA}}
received the B.S. degree in computer science from Ateneo de Manila University, NCR, Philippines in 2016. He is currently pursuing a Ph.D. degree in computer science in Fordham University, NY, USA.

From 2016 to 2022, he worked as a Data Analyst and a Cloud Solutions Developer in industry before pursuing graduate studies. His research interests include the use of classical systems to support the operation of quantum computers and the application of distributed systems principles into quantum architectures.
\end{IEEEbiography}

\vskip -2.5\baselineskip plus -1fil
\begin{IEEEbiography}[{\includegraphics[width=1in,height=1.25in,clip,keepaspectratio]{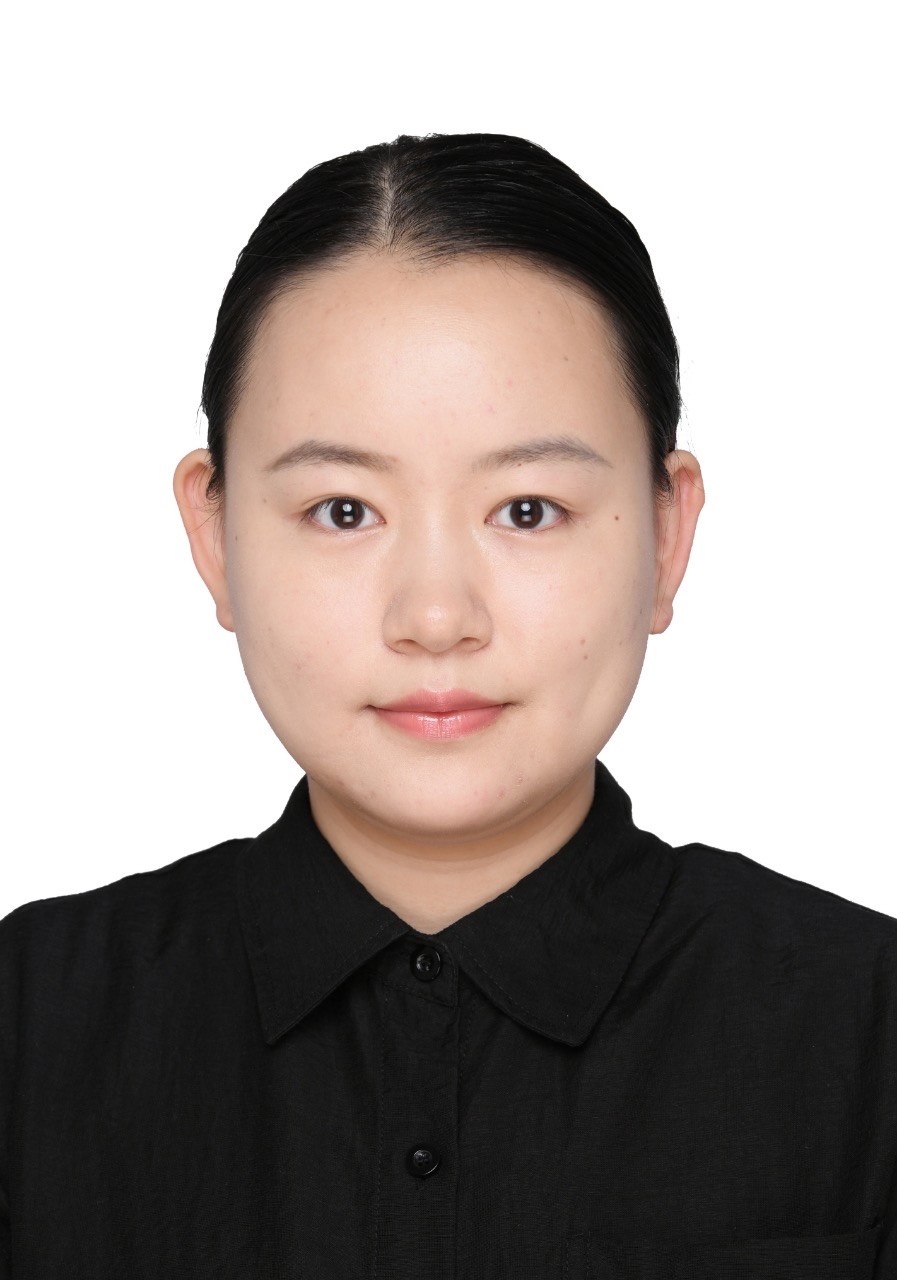}}]%
{\textbf{SHUWEN KAN}} is a Ph.D. student in Computer Science at the Department of Computer and Information Science, Fordham University. She received her M.S. in Data Science from Johns Hopkins University in 2023 and her B.S. in Applied Mathematics and Statistics from Purdue University in 2021. Her research interests lie at the intersection of quantum computing systems, compiler design, and fault-tolerant quantum computing, focusing on developing scalable architectures for heterogeneous quantum systems and leveraging machine learning tools for system optimization and design.
 \end{IEEEbiography}

\vskip -2.5\baselineskip plus -1fil
\begin{IEEEbiography}[{\includegraphics[width=1in,height=1.25in,clip,keepaspectratio]{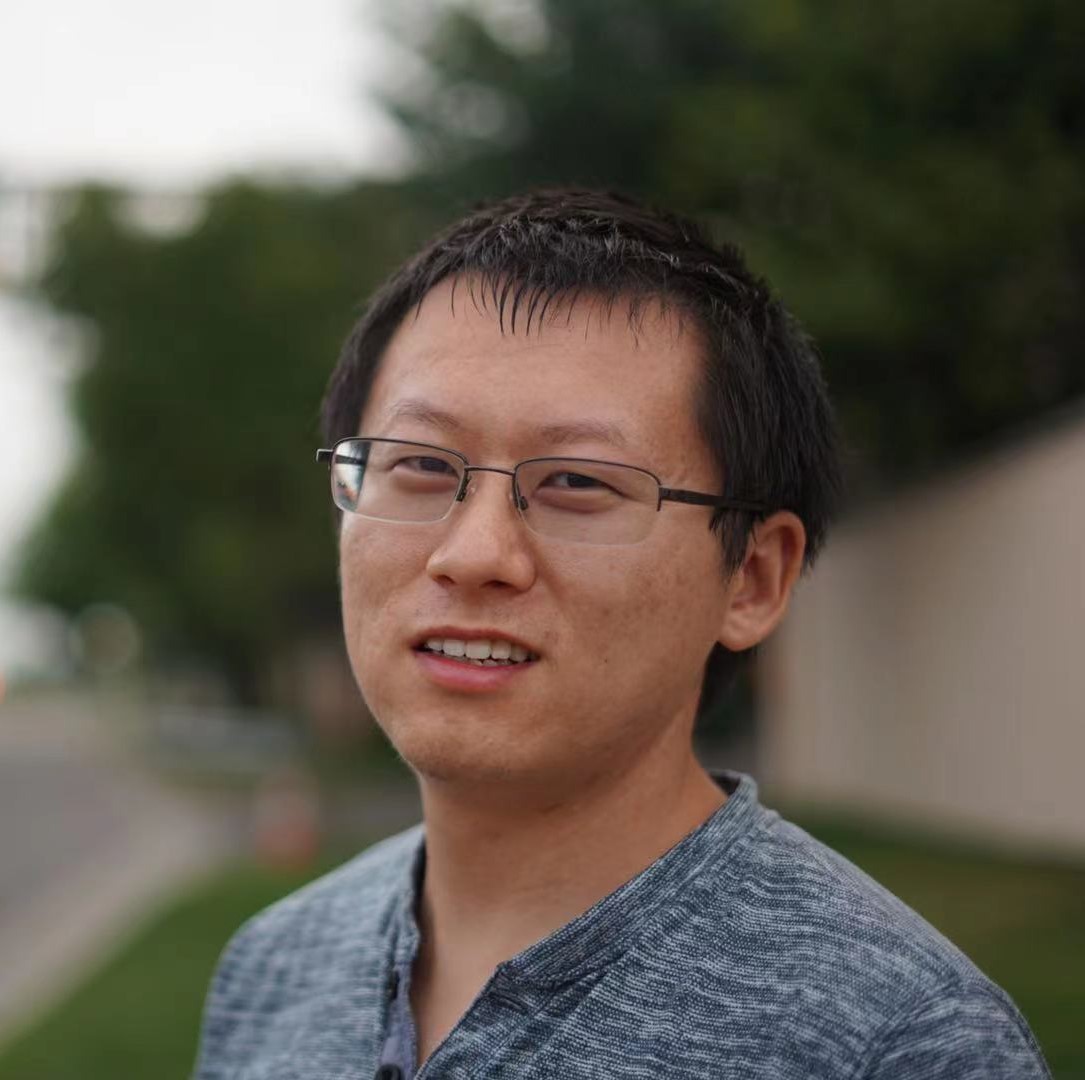}}]%
{\textbf{WENQI WEI}}
is a tenure-track assistant professor in the Department of Computer and Information Sciences at Fordham University. He 
obtained his PhD in the School of Computer Science, Georgia Institute of Technology in 2022. 
His research interests include trustworthy
AI, data privacy, machine learning service, and big
data analytics. His research has appeared on major cybersecurity, data mining, and AI venues, including CCS, CVPR, IJCAI, theWebConf, IEEE TDSC, IEEE TIFS, and ACM CSUR. He is an associate editor of ACM Transactions on Internet Technology and IEEE Transactions on Big Data.
\end{IEEEbiography}

\vskip -2.5\baselineskip plus -1fil
\begin{IEEEbiography}[{\includegraphics[width=1in,height=1.25in,clip,keepaspectratio]{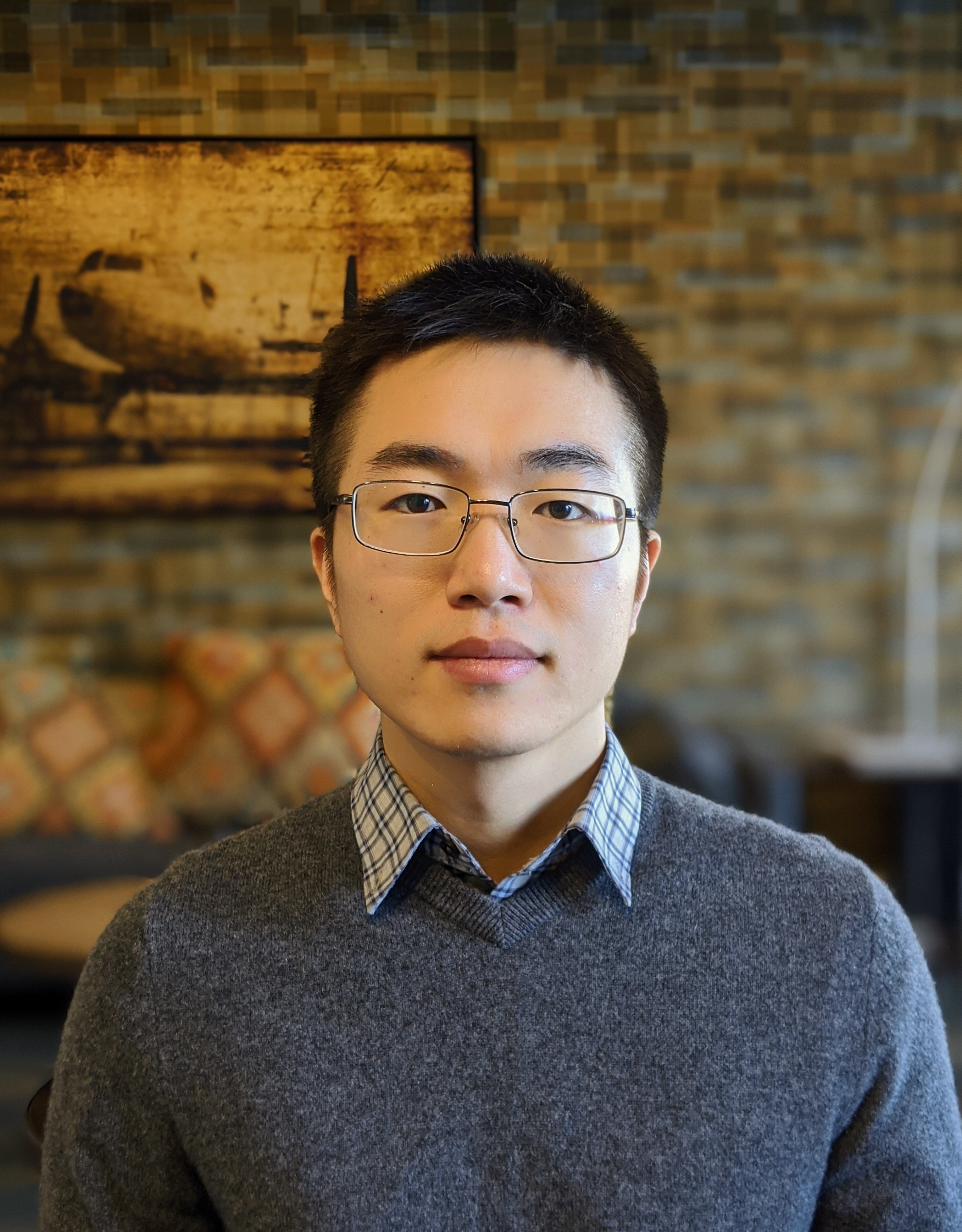}}]%
{\textbf{JUNTAO CHEN}}
is an Assistant Professor at the Department of Computer and Information Sciences, Fordham University. He received his Ph.D. degree in electrical engineering from New York University in 2020. His research interests include cyber-physical security and resilience, quantum AI and quantum computing security, and game and decision theory. 
\end{IEEEbiography}

\vskip -2.5\baselineskip plus -1fil
\begin{IEEEbiography}[{\includegraphics[width=1in,height=1.25in,clip,keepaspectratio]{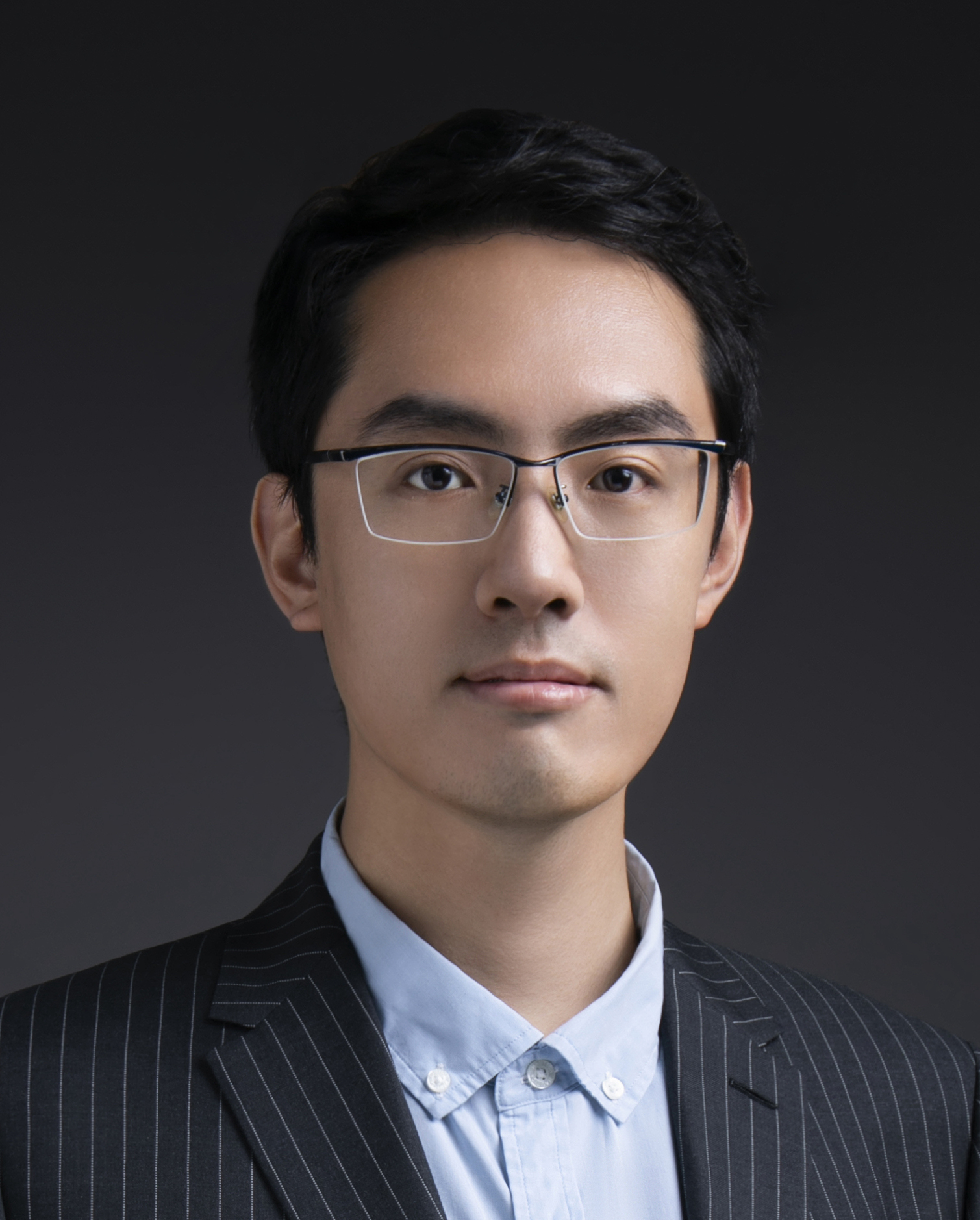}}]%
{\textbf{KAIXUN HUA}}
is an Assistant Professor at the Department of Industrial and Management Systems Engineering, University of South Florida. He was a Postdoctoral Fellow at the Institute of Applied Mathematics and Department of Chemical and Biological Engineering, University of British Columbia. He obtained his Ph.D. degree from the University of Massachusetts Boston in 2019, an M.Eng. degree from Cornell University in 2013 and a B.S. degree from Shanghai Jiao Tong University in 2012. His research focuses on developing scalable deterministic global optimization algorithms to address challenges of interpretable machine learning systems. 
\end{IEEEbiography}

\vskip -2.5\baselineskip plus -1fil
\begin{IEEEbiography}[{\includegraphics[width=1in,height=1.25in,clip,keepaspectratio]{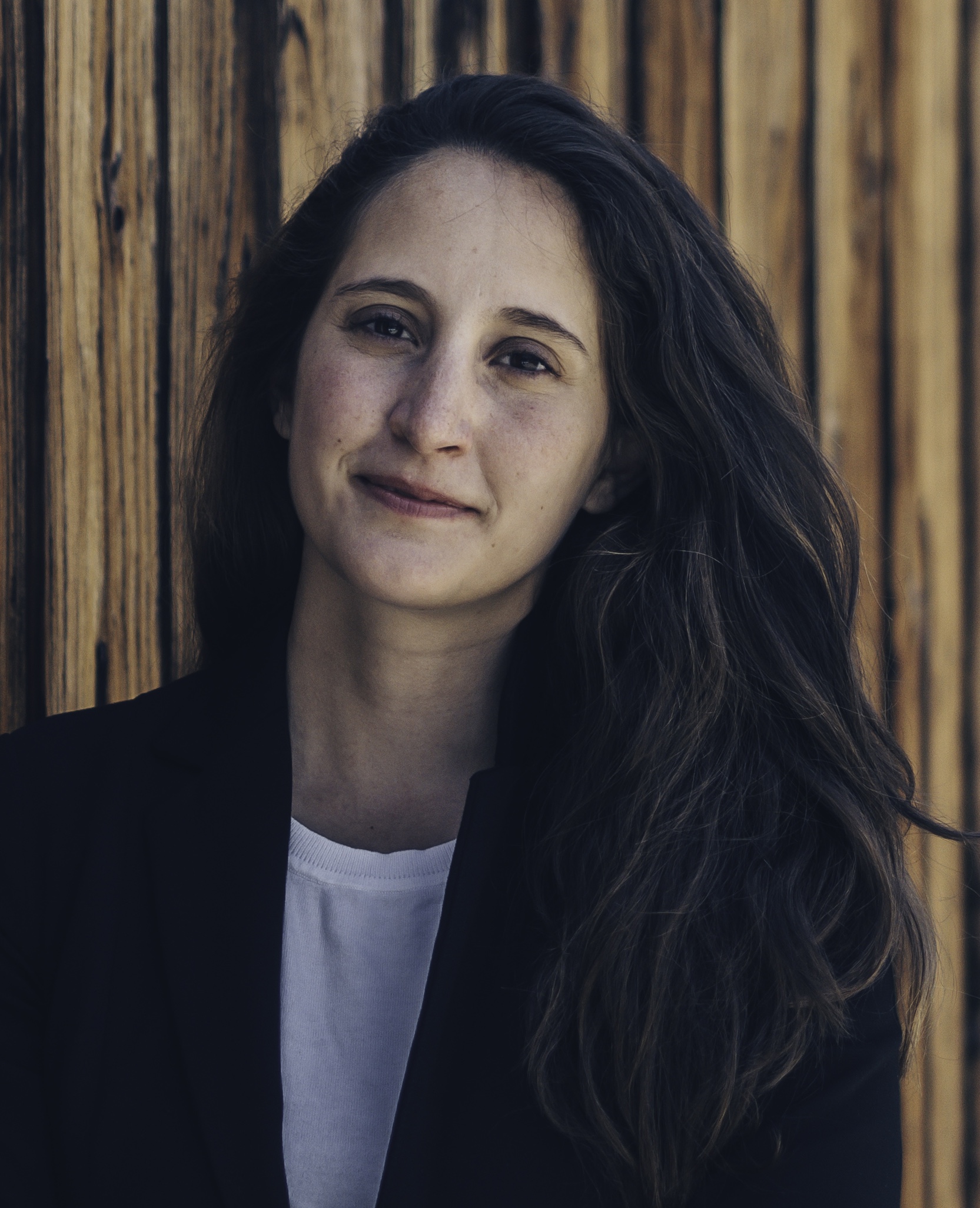}}]%
{\textbf{SARA MOURADIAN}} received her B.S., M.Eng., and Ph.D. degrees from the Massachusetts Institute of Technology in Electrical Engineering and Computer Science in 2010, 2012, and 2018, respectively. She was then a Intelligence Community Postdoctoral Fellow in the University of California, Berkeley Physics department.

She is currently an Assistant Professor of Electrical and Computer Engineering at the University of Washington, Seattle. Her research focuses on enabling utility-scale quantum technologies through optical engineering, electrical engineering, and new control, sensing, and calibration techniques
\end{IEEEbiography}

\vskip -2.5\baselineskip plus -1fil
\begin{IEEEbiography}[{\includegraphics[width=1in,height=1.25in,clip,keepaspectratio]{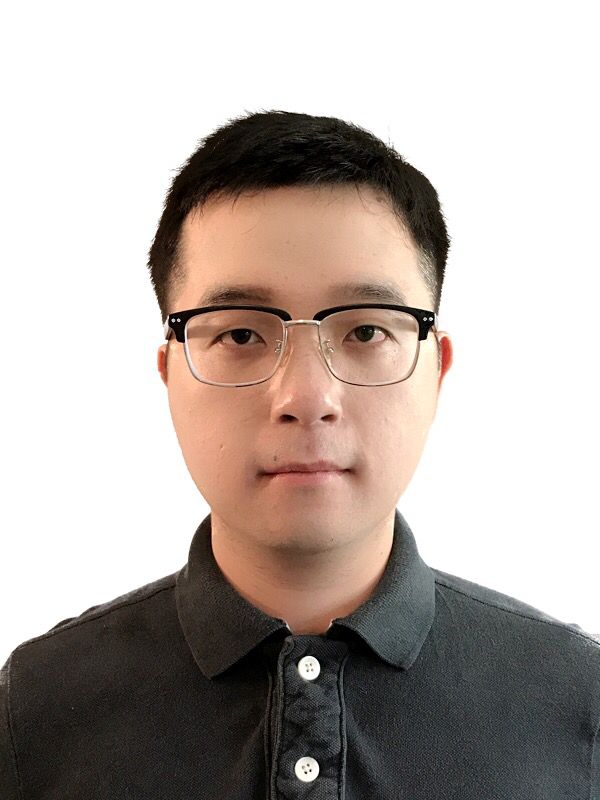}}]%
{\textbf{YING MAO}}
is an Associate Professor in the Department of Computer and Information Science, Fordham University, New York, NY, USA. He received the Ph.D. degree in Computer Science from the University of Massachusetts Boston, Boston, MA, USA in 2016. He was previously a Fordham-IBM Research Fellow. His research interests include quantum systems, quantum machine learning, quantum-classical optimization, quantum system virtualization, cloud resource management, data-intensive platforms, and containerized applications.
\end{IEEEbiography}

\EOD
\end{document}